\documentclass{aastex}
\usepackage{spr-astr-addons}
\usepackage{url}

\RequirePackage{color}

\begin{document}

\title{Effective recombination coefficient and solar zenith angle effects on low-latitude D-region ionosphere evaluated from VLF signal amplitude and its time delay during X-ray solar flares}
\slugcomment{Not to appear in Nonlearned J., 45.}
\shorttitle{Short article title}
\shortauthors{Autors et al.}

\author{Tamal Basak} \and \author{Sandip K. Chakrabarti\altaffilmark{1}}
\affil{S. N. Bose National Centre for Basic Sciences, Block-JD, Sector-III, Salt Lake, Kolkata- 700098, India}

\altaffiltext{1}{Indian Centre for Space Physics, 43 Chalantika, Garia Station Road, Garia, Kolkata- 700084, India}

\begin{abstract}
Excess solar X-ray radiation during solar flares causes an enhancement of ionization 
in the ionospheric D-region and hence affects sub-ionospherically propagating VLF signal 
amplitude and phase. VLF signal amplitude perturbation ($\bigtriangleup A$) and amplitude time delay ($\bigtriangleup t$)
(vis-\'a-vis corresponding X-ray light curve as measured by GOES-15) of NWC/19.8 kHz 
signal have been computed for solar flares which is detected by us during Jan-Sep 2011.
The signal is recorded by SoftPAL facility of IERC/ICSP, Sitapur
(22$^\circ$ 27$^{\prime}$N, 87$^\circ$ 45$^{\prime}$E), West Bengal, India. In first part 
of the work, using the well known LWPC technique, we simulated the flare induced
excess lower ionospheric electron density by amplitude perturbation method.
 Unperturbed D-region electron density is also obtained from simulation and compared with IRI-model results.
Using these simulation results and time delay as key parameters, we calculate
the effective electron recombination coefficient ($\alpha_{eff}$) at solar flare peak region. 
Our results match with the same obtained by other established models. In 
the second part, we dealt with the solar zenith angle effect on D-region during
 flares. We relate this VLF data with the solar X-ray data. 
We find that the peak of the VLF amplitude occurs later than the time of the X-ray peak 
for each flare. We investigate this so-called time delay ($\bigtriangleup t$). 
For the C-class flares we find that there is a direct 
correspondence between $\bigtriangleup t$ of a solar flare and the average solar zenith angle $Z$
over the signal propagation path at flare occurrence time. Now for deeper 
analysis, we compute the $\bigtriangleup t$ for different local diurnal time slots $DT$. We find that 
while the time delay is anti-correlated with the flare peak energy flux $\phi_{max}$ independent of
these time slots, the goodness of fit, as measured by $reduced$-$\chi^2$, actually worsens as the day progresses.
The variation of the $Z$ dependence of  $reduced$-$\chi^2$ seems to 
follow the variation of standard deviation of $Z$ along the $T_x$-$R_x$
propagation path. In other words, for the flares having almost constant $Z$ over the path 
a tighter anti-correlation between $\bigtriangleup t$ and $\phi_{max}$ was observed. 
\end{abstract}

\keywords{X-ray solar flare, time delay, recombination coefficients, D-region ionosphere}

\section{Introduction}

The ionosphere of the earth is a gigantic detector and it characteristically 
responds to the ionizing agents of terrestrial and extra-terrestrial origin. In particular,
it is long known that the hard and soft X-rays originating from solar flares (Mitra 1974; Pant 1993)
and strong compact celestial sources like X-ray novae, galactic centre do perturb
the lower ionospheric D-region (Sharma et al. 1972; Kasturirangan et al. 1976). 
The excess EUV and X-ray 
radiations produced during the solar flares cause excess ionization 
\citep{basa10,garci07,chak12}. This manifests in an enhancement of the 
free electron density of the D-region by several orders of magnitude 
than the normal value. This is primarily due to the electron detachment from 
nitrogen and oxygen \citep{thom01}. The main physical mechanism behind 
the ionospheric effects of the solar flares have been discussed by 
\citet{mitra74}. A comparative study between the changes of VLF signal 
amplitude (and phase) and solar X-ray flux has been made by several 
workers such as \citet{anan73} and \citet{pant93}. 

Through some recombining
electron loss processes, mainly, the electron-ion recombination, ion-ion recombination and 
three body recombination, the ionosphere gradually comes back to its normal condition.
In the present paper, we analyse VLF signal data coming from NWC/19.8 kHz and recorded at 
Ionospheric and Earthquake Research Centre (IERC) of Indian Centre for Space Physics (ICSP), 
at Sitapur (22$^\circ$ 27$^{\prime}$N, 87$^\circ$ 45$^{\prime}$E), India. 
The $T_x$-$R_x$ $Great$ $Circular$ $Path$ is $5691 kms$ long and average signal attenuation
along this path is low. We present the VLF response along with the corresponding X-ray flux of the solar flares which were detected during 
Jan-Sep 2011 and studied the evolution of recombination of the D-region during several
classes of flares. VLF amplitude time delay ($\bigtriangleup t$) and electron 
density ($N_e$) are used as key parameters to obtain the recombination  coefficient $\alpha_{eff}$.

To handle D-region chemistry at lower heights ($\sim$ $70$ km) \citet{mitra72} prescribed
 a 6-ion + electron scheme, where the ions are O$_2$$^-$, NO$^+$, O$_2$$^+$, O$_4$$^+$,
 H$^+$(H$_2$O)$_n$ and NO$_3$$^-$(H$_2$O)$_n$. After studying all reaction channels
 \citet{mitra72} showed that, at those lower heights, even due to significant flare induced
 ionization, the variation of positive and hydrated ions are less. In
 D-region above $74$-$75$ kms the negative ions are practically absent. There
 the main ion contents are NO$^+$ and H$_2$O$_5$$^+$. At $75$ to $80$ km height,
 the production of O$_2$$^+$ is less than NO$^+$ under quiet and flare X-ray conditions 
\citep{mitrarowe74}. So main clustering is done with NO$^+$, through
 NO$^+$$\rightarrow$NO$^+$.CO$_2$$\rightarrow$H$^+$(H$_2$O)$_3$ \citep{mitra72}. 
Clustering of O$_2$$^+$ and O$_4$$^+$ are possible only during direct photo-ionization
 process under flare conditions. But a large fraction of hydrates are lost as they
 go back to O$_2$$^+$ and NO$_2$$^+$. Thus under flare conditions [H$_2$O] and [O] 
varies, and as ionization rates go up the percentage of hydrate ion goes down with
 $\alpha_{eff}$. For IERC-NWC low-latitude propagation path this chemistry is significant even
 at 74 km as it is evident from our results regarding $\alpha_{eff}$. 

Early studies of the recombination of electrons and ions in lower ionosphere
were made by \citet{appl53,mitra53,mitra74}. Subsequently, in a series of papers,
from the ionospheric data obtained from rocket measurements and other experiments and using
the electron continuity theory, the dissociative recombination coefficient ($\alpha{_D}$), mutual ionic recombination
coefficient ($\alpha{_i}$), electron to negative ion ratio ($\lambda$) 
etc. are estimated and modelled \citep{whit61,pop62,whit62}.
\citet{mitra63} has reported different methods to estimate recombination
coefficients (including height profile) at different ionospheric conditions, e.g.,
diurnal asymmetry method, eclipse method and solar decay curve method. Values 
of $\alpha{_i}$ and $\alpha{_D}$ obtained by the above mentioned methods agree with the experimental
results. \citet{whit65,par65,gled86} have developed theoretical methods for empirical
expressions for $\alpha_{eff}$ (it includes $\alpha{_i}$, $\alpha{_D}$, $\lambda$ and
other dust capture coefficients of aerosol theory). \citet{wag72} proposed a different
method namely `Thomson Scatter Experimental Technique' to simulate electron density
and $\alpha_{eff}$ during solar flares relative to pre-flare condition at E-region. According
to \citet{appl53}, the $N_e$ and $\alpha_{eff}$ are directly related to time delay $\bigtriangleup t$.
This $\bigtriangleup t$ appears due to `inertial' properties and chemical reaction time scales
of the D-region ionosphere. This time delay is defined as,
\begin{equation}
{\bigtriangleup t} = APT - FPT,
\end{equation} 
where, $APT$ is the Amplitude Peak Time of VLF signal and $FPT$ is 
the X-ray Flare Peak Time. This $\bigtriangleup t$ is nearly similar to the $sluggishness$
\citep{appl53,valni72} and $relaxation$ $time$ \citep{mitra74}. \citet{bal91} has analysed
photo-ionization rates in detail for different ion constituents (NO$^+$,O$_2$$^+$) and for
different solar X-ray bands. Hence assuming photo-chemical equilibrium, the height profile
of $\alpha_{eff}$ for different M and X classes of flares has been computed. In last few
years, the calculation of the recombination coefficients for different types of solar and
geomagnetic events have been done. \citet{poz97} estimated $\alpha_{eff}$ of nighttime
auroral ionosphere using EISCAT radar data and showed clearly that $\alpha_{eff}$ is
greatly decreased with the increase in particle precipitation flux at higher energy. The same
EISCAT radar observation is used by \citet{ose09} to model $\alpha_{eff}$ during solar
proton events. To compare experimental measurements of $\alpha_{eff}$ with those obtained from the theory
\citet{fri04} inserted temperature and pressure of lower ionosphere in $\alpha_{eff}$
expression empirically. Using two different approaches of D-region electron
continuity theory, workers such as \citet{zig07} calculated $\alpha_{eff}$ during the flare peak
and \citet{nina11} during the decay regime of a flare. 

In the later section of the paper we analysed the zenith angle ($Z$) dependence 
of $\bigtriangleup t$ in greater detail.  First, we compute the time delay of 
the response of the VLF signals with respect to all types of 
classes of solar X-ray flux variation. 
For NWC-IERC path, positive VLF amplitude change is observed 
for flares considered here. Similar results are also reported 
for NAA/24.0 kHz to Belgrade path by \citet{zig07}. 
The $\bigtriangleup$t measured for NWC-IERC path is found to 
be positive. We showed that this $\bigtriangleup$t depends on the 
solar zenith angle at flare occurrence time along the signal propagation path.

There are several works in the literature on the changes in VLF signal amplitude and phase 
and the time delay have been carried out for studying D-region evolution due to solar flares. 
Earlier theoretical works on ionospheric changes were done 
by \citet{wait62}, \citet{wait64}, \citet{mitra74} etc. 
\citet{thom01},\citet{mcrae04} and \citet{thom05} probed the D-region changes 
during flares through the VLF amplitude analysis. 
In a review, \citet{tsu09} discussed the long term effects 
on the ionosphere by solar flares followed by Fast 
Interplanetary Coronal Mass Ejections (ICMEs). \citet{qian11} showed 
the dependency of TEC of ionosphere on solar zenith 
angle. \citet{le07}, \citet{zhang11}, \citet{le12} and others 
reported that flare induced Total Electron Content (TEC) 
decreases when zenith angle increases. All these works 
discuss the effects of solar zenithal angle on upper and 
middle ionosphere. 

Mitra (1974) established a relation of $\tau$ with the
 maximum electron density ($N_{e,max}$). Mitra (1974) showed that
during a given solar flare, the `$\tau$' is inversely related with $N_{e,max}$, i.e., 
for stronger flares, the ionospheric response is more instantaneous.
Similar kind of results is reported by Valnicek and Ranzinger (1972). Using the X-ray data obtained by Inter-cosmos 1 satellite, 
they showed that the `sluggishness' ($\bigtriangleup t$) decreases when
solar induced ionizing activity increases and vice versa. Zigman et al. (2007) calculated
the time variation of $N_e$ during several classes of flares using `electron
continuity theory' and $\bigtriangleup t$ as crucial parameter (where, VLF signal
along NAA/24.0 kHz to Belgrade propagation baseline are used for analysis).
\citet{thom01} studied both long (NLK/24.85 kHz to Dunedin, $12.3 Mm$) and short
 (French T$_x$/18.3 kHz to Cambridge, $617 km$) path VLF propagation. They showed
 that for VLF amplitude perturbation due to wide variation of $Z$ over the path the $Z$-effect is less for such a long path and oppositely
 $Z$-effect is significant for shorter path. \citet{gru05} monitored the VLF
 response during flares for a shorter path (GQD/22.1 kHz to Belgrade,$\sim$
 $2000 km$) and they showed that the $Z$-effect on amplitude and phase delay
 are not prominent enough but the flares occurred at higher zenith angle (i.e
. during dusk and dawn) can cause amplitude and phase delay if they are strong
 enough. But in this paper we will analyse $\bigtriangleup t$ and its $Z$-effect.

In this paper, we firstly formulate a method to estimate the {\it effective recombination coefficient} 
($\alpha_{eff}$) at a low-latitude D-region ionospheric with the help of some  
established methods mentioned in the earlier paragraphs, where, the VLF amplitude time
delay ($\bigtriangleup t$) and VLF amplitude perturbation ($\bigtriangleup A$) during
a solar flare are crucial parameters. We particularly concentrated to  calculate electron density because
 the positive and negative ions, due to their comparatively heavier masses than electron,
 hardly affect the VLF propagation in lower ionosphere \citep{mitra92}. In the next section, we describe the way we estimate
$\bigtriangleup t$ and $\bigtriangleup A$ from our VLF data.
Using the well known Long Wave Propagation Capability (LWPC) code \citep{ferg98}, we simulate
the maximum electron density ($N_{e,max}$) at the flare peak, where $\bigtriangleup A$$_{max}$
has been used as the input parameter. The `Range Exponential' sub-program of LWPC which follows two parameter
($h^\prime$, $\beta$) model of lower ionosphere \citep{wait64}, has been used for simulation.
Finally, we use the electron continuity theory \citep{appl53,mitra63,mitra74,zig07} and
calculate $\alpha_{eff}$ for all classes of flares. A good correlation has been
found between $\alpha_{eff}$ and the peak flux of solar flare ($\phi_{max}$). Moreover, the obtained
values of $\alpha_{eff}$  by our method appear to be close to the values presented in the literature.

The second part of the work consists of two different analysis. First of which
 is related to $Z$, we studied how $\bigtriangleup t$s measured
for C-class flares depend on $Z$ computed along the propagation path during the flare occurrence time. 
We investigated a total of $22$ flares. As the $\bigtriangleup t$
has been found to depend on the strength of flares, we chose only the C-class 
flares. We observe that the average zenith angle (Z) over 
the path during flare occurrence has a linear correlation with $\bigtriangleup$t. 
We compute this correlation quantitatively. Ours is an alternative 
method to estimate the relation between $N_e$-profile 
of D-region and $Z$, because $\bigtriangleup$t is directly connected to $N_e$ \citep{mitra74,zig07}. 
Similar correlations of time delay with Z can be tested for the M-class or X-class flares 
but since their numbers are much lower, for a good statistics, we need to collect the data
for a longer period. Now in the later part for deeper analysis, we chose
78 flares of C, M and X-classes which occurred at different diurnal
times. We grouped them into six separate equal sized time bins ($DT$) according to their
occurrence times. For each bin, we fitted $\bigtriangleup t$ versus peak flare
flux ($\phi_{max}$) of respective flares with an empirically chosen exponentially decaying
function and calculated $reduced$-$\chi^2$ to estimate the goodness of fit. Finally,
we compute the average and standard deviation ($\sigma$) of the solar zenith angle $Z$ 
(averaged over the $T_x$-$R_x$ propagation path) for each $DT$, and 
the $reduced$-$\chi^2$ varies exactly in the same way as
the standard deviation ($\sigma$) of zenith angle ($Z$) distributed over $T_x$-$R_x$
path within each time bin $DT$.

\section{Monitoring of solar flares using VLF radio signal}

For VLF amplitude measurement in this paper, we record the electric field component of
 NWC (19.8 kHz) transmitter signal in the units of dB above
 1$\mu$V m$^{-1}$ and with $1$ $s$ resolution. Besides, the VLF signal
 amplitude and phase from VTX, JJI and other transmitters are also recorded wherever available. The recording
 has been done by SoftPAL, a fully software version of AbsPAL (Absolute Phase and
 Amplitude Logger) - developed by the Radio and Space Physics Group of Otago
 University, New Zealand. During solar flare all VLF data are compared with solar X-ray data taken in the $0.1 - 0.8$ nm
($\phi$ in W m$^{-1}$) obtained from GOES-15, National Oceanic and Atmospheric
 Administration (NOAA), USA. The recording period of data in this paper belongs to the rising
 phase of Solar Cycle no. 24. We got $\bigtriangleup A$, $\bigtriangleup t$ $\textgreater$0
 for all the recorded cases (see Table 1). Similar results are reported by \citet{thom01} for NLK/24.8 kHz
 to Dunedin, New Zealand path. Though we detected hundreds of flares during
 this period, but for $\alpha_{eff}$ analysis we sorted out $22$ flares (C \& M-classes) which
 occurred close to midday ($Z$ mostly within $15^\circ$ to $30^\circ$, 
see Table 1). This is because $\bigtriangleup A$, $\bigtriangleup t$ and $N_e$ have
significant dependence on $Z$ averaged over the signal
 propagation path at occurrence time \citep{han11}. Thus we eliminated the
 major effects of $Z$ and showed that $\alpha_{eff}$ has an inverse relation
with $\phi_{max}$. Though the ionospheric relaxation time ($\tau$) is an intrinsic characteristics of 
D-region but $\alpha_{eff}$ varies with electron production rate ($q$). Different $q$ values stand for
different flare strengths. Through rigorous analysis \citet{mitra72} showed that as $q$ goes
 higher, then percentage of hydrated cluster ions in the D-region gets lowered at
 faster rate and percentage of simple positive ions like [O] gets higher. 
Thus complex ion chemistry reduces to simpler molecular ion chemistry and
 as the consequence of that $\alpha_{eff}$ get reduced. 
\begin{figure}[h!]
\begin{center}
\includegraphics[height=6.2cm]{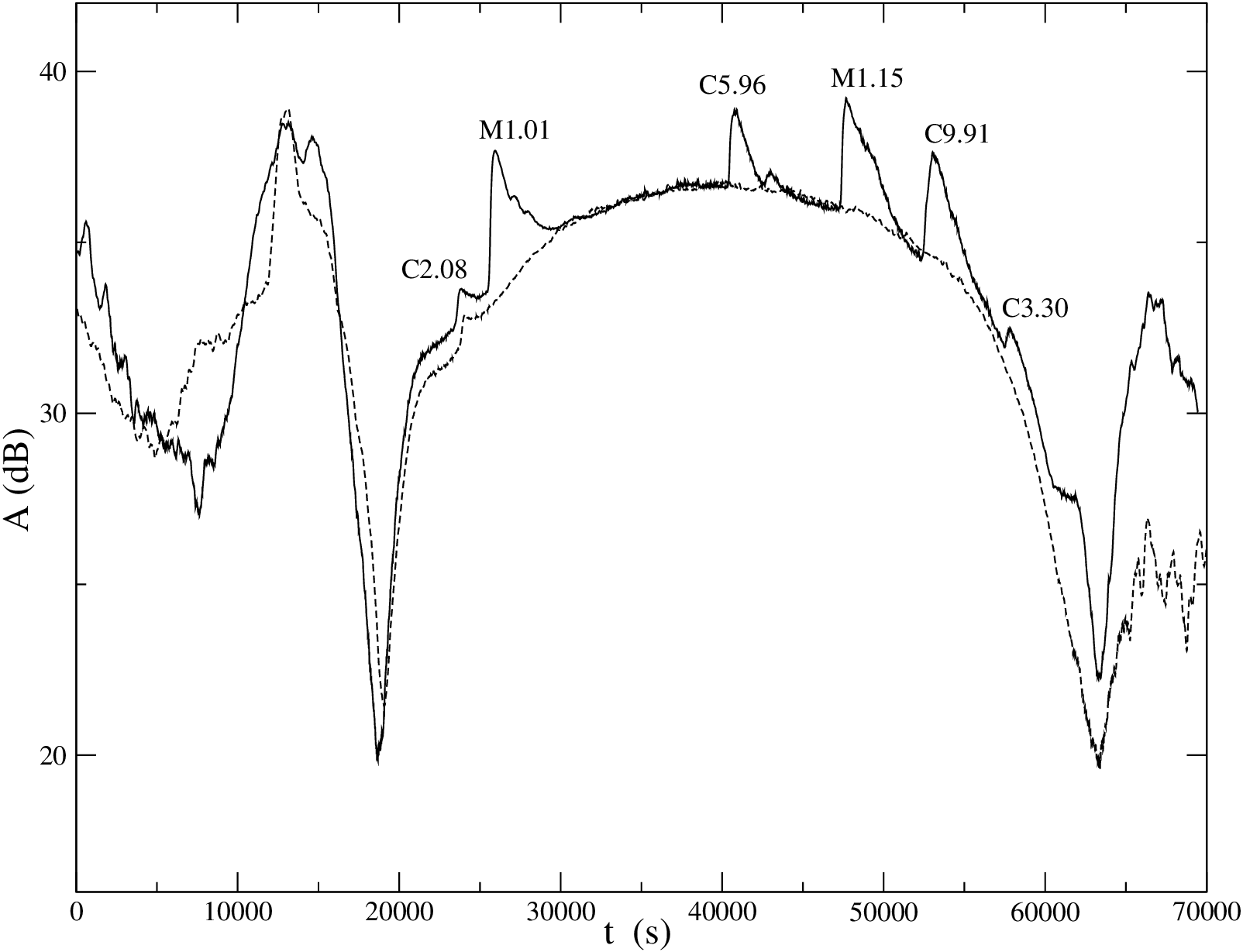}
\end{center}
\caption{The amplitude A (left) and phase (right) of VLF signals transmitted 
from NWC (19.8 kHz), as a function of time as recorded at IERC(ICSP), Sitapur 
(22$^\circ$ 27$^{\prime}$N, 87$^\circ$ 45$^{\prime}$E) on 16th Feb (solid line) 
and 12th Feb (dotted line) 2011. Six solar flares of different classes (C2.08 
at 06:37 hrs, M1.01 at 07:11 hrs, C5.96 at 11:19 hrs, M1.15 at 13:14 hrs, 
C9.91 at 14:43 hrs and C3.3 at 16:04 hrs) were recorded on 16th Feb 2011. 
All times are in IST(=UT+5.5 hrs)} 
\end{figure}

In Fig. 1, we present a typical diurnal variation of 19.8 kHz VLF 
signals recorded on the 16th Feb 2011 (a solar-active day with 
several solar flares at different times) and 12th Feb 2011 
(a solar-quiet day). A total six solar flares of different 
classes are detected on the 16th Feb 2011. The sharp rise 
and the slow decay pattern of a typical X-ray irradiance is 
exhibited by the VLF signal amplitude and phase in these observations. 
\begin{figure}[h!]
\begin{center}
\includegraphics[height=6.5cm]{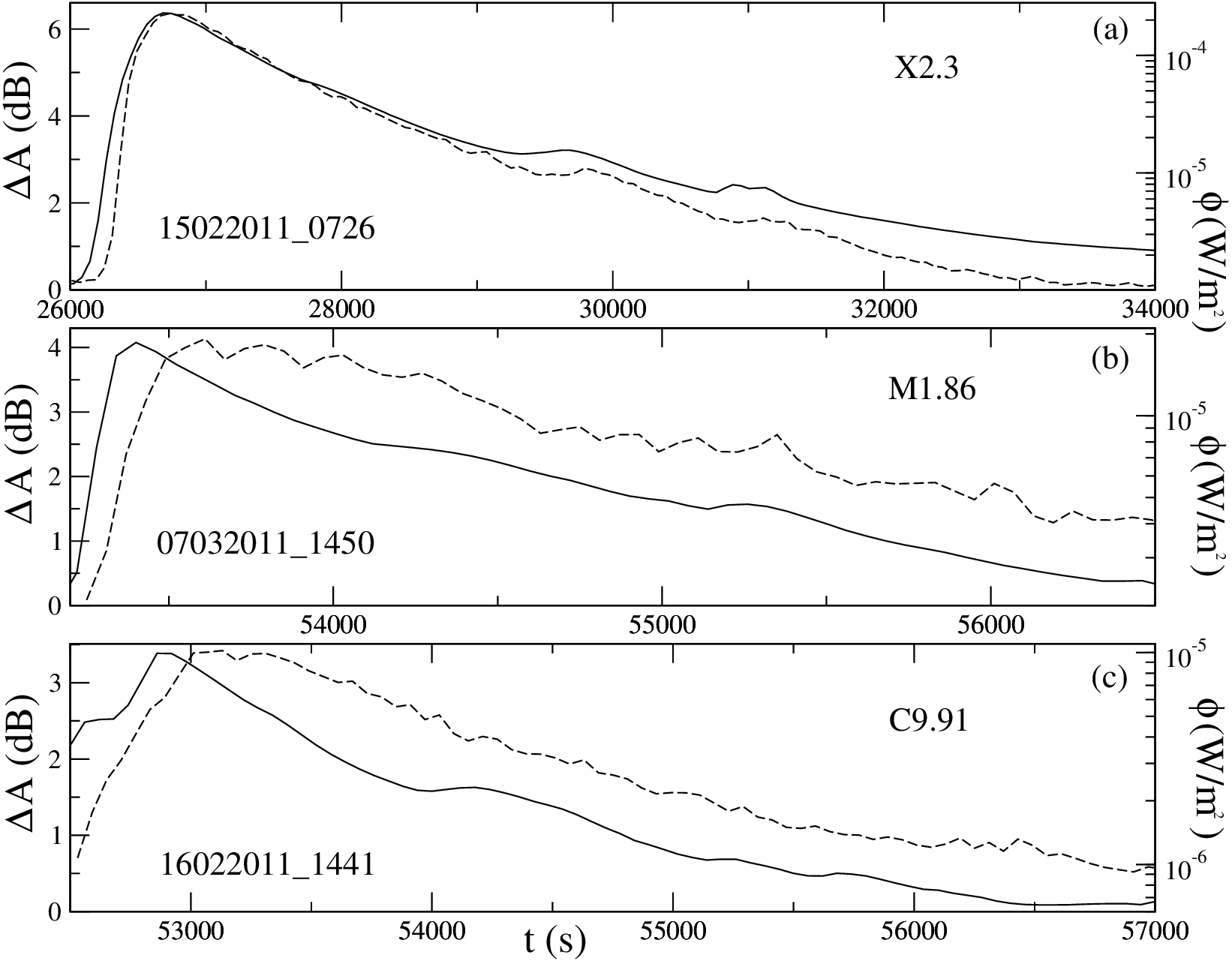}
\end{center}
\caption{Examples of (a) X-class (b) M-class and (c) C-class solar
 X-ray flares as recorded by GOES-15(solid lines) along with
 corresponding VLF signal amplitude deviations ($\bigtriangleup A$) (dashed lines).
 Date and time (DDMMYYYY$\textunderscore$HHMM) are marked in respective
panels.}
\end{figure}
In Fig. 2, we present a few examples of typical VLF amplitude response for X, M and C class flares in our 
receiver. These are superposed on the X-ray light curve (details of those flares are
 marked in the Figure). The general pattern of the X-ray
 light curve is same as the VLF amplitude.

Now to check the direct correspondence between $\bigtriangleup t$ and $Z$ we sorted 
another $22$ C-class flares. A total of 78 flares are of all classes are considered
 to estimate $Z$ dependence of $reduced$-$\chi^2$.

\section{Estimation of $\bigtriangleup t$ and $\bigtriangleup A$$_{max}$}

Maximum perturbed VLF amplitude ($\bigtriangleup A$$_{max}$) and VLF amplitude 
time delay ($\bigtriangleup t$) are important parameters as mentioned in the
Sec. 1. These are defined as,
\begin{equation}
{\bigtriangleup A_{max}} = A_{perturbed,max} - A_{quiet},
\end{equation}
where, $A_{perturbed, max}$ is maximum VLF perturbation for a given solar flare
and $A_{quiet}$ is the mean of the available 5 solar-quiet day data which are
closest to the `flare day'. We have chosen solar-quiet days, such that they are free 
from even smaller perturbation  due to tiny flares. $\bigtriangleup A$$_{max}$ is in dB units. Now,
\begin{equation}
{\bigtriangleup t} = t_{\bigtriangleup A_{max}} - t_{\phi_{max}},
\end{equation}
where, $t_{\bigtriangleup A_{max}}$ is the time of $\bigtriangleup A$$_{max}$ and
 $t_{\phi_{max}}$ is the time of peak flux ($\phi_{max}$) of solar flare.  Fig. 3
 shows $\bigtriangleup t$ schematically. We fit
 the peak regions of $\bigtriangleup A$ and $\phi$ data sets with higher order
polynomials
 and obtained the times at which the peaks occur. These are called $t_{\bigtriangleup
A_{max}}$
 and $t_{\phi_{max}}$ respectively. Overall this method reduces the error drastically.  Most importantly,
this fitting 
 is important for correct estimation of $t_{\phi_{max}}$ as the maximum
resolution of
 online available data is one minute.

This fitting method is repeated for all 
the flares analysed in this paper. These $\bigtriangleup A$$_{max}$ and $\bigtriangleup t$
are shown in Table 1 and also plotted as functions of respective $\phi_{max}$ in Fig. 4 and Fig. 5. In Fig. 5, we
note that $\bigtriangleup t$ has a tendency of decreasing with increasing $\phi_{max}$.
We also find that $\bigtriangleup A$$_{max}$
for $\sim$ M9.0 class flare has gone up to $\sim$ 4 dB and $\bigtriangleup t$ went
down to $\sim$ $70$ s (see Table 1).

\begin{figure}[h!]
\begin{center}
\includegraphics[height=6.4cm]{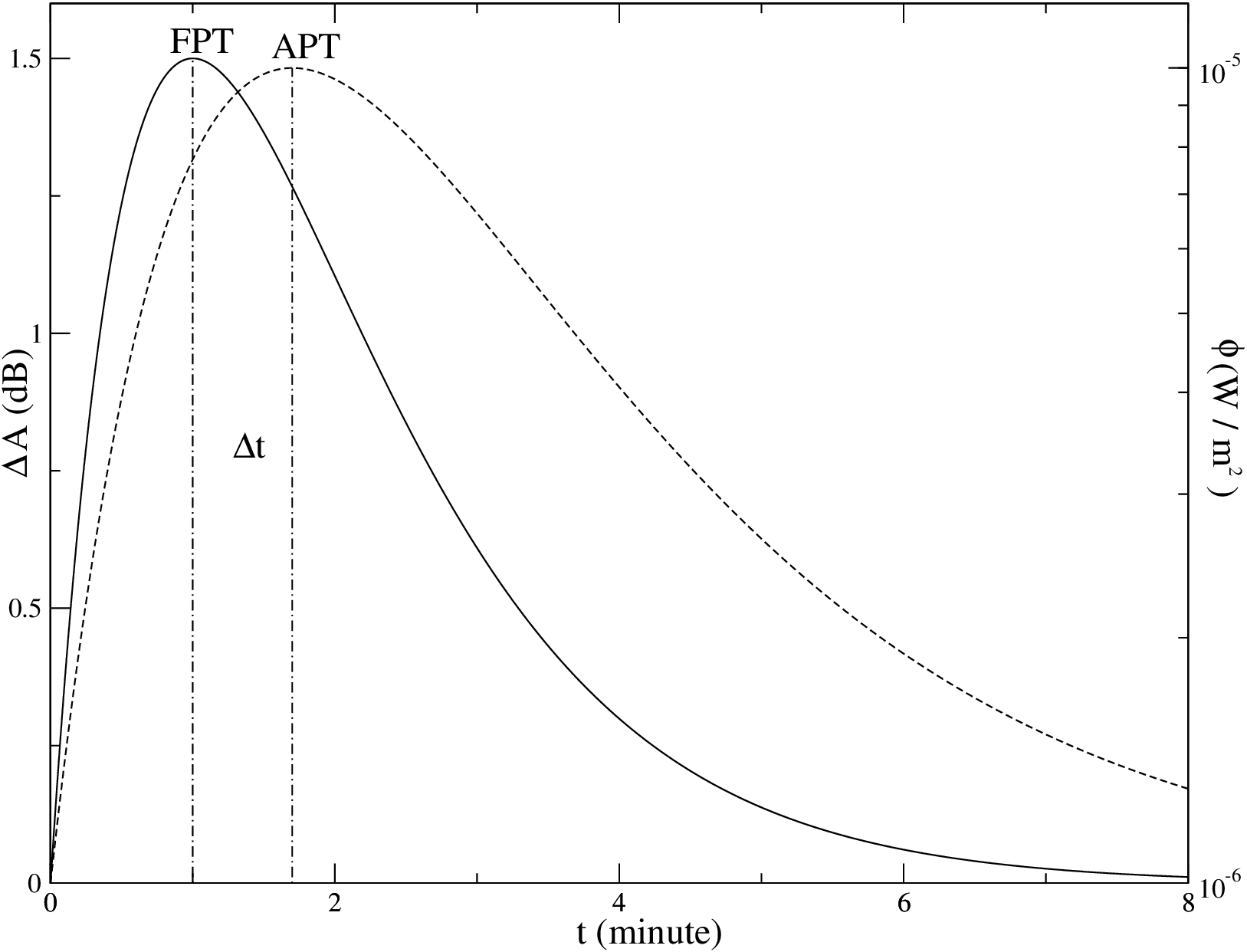}
\end{center}
\caption{A schematic diagram showing the meaning of $\bigtriangleup$t, 
the time delay of the VLF signal. The solid curve represents a typical 
soft X-ray irradiance $\phi$ (W/m$^{2}$) and the dashed curve 
represents the corresponding VLF amplitude deviation 
($\bigtriangleup$A) in dB }
\end{figure}

\begin{figure}[h!]
\begin{center}
\includegraphics[height=6.3cm]{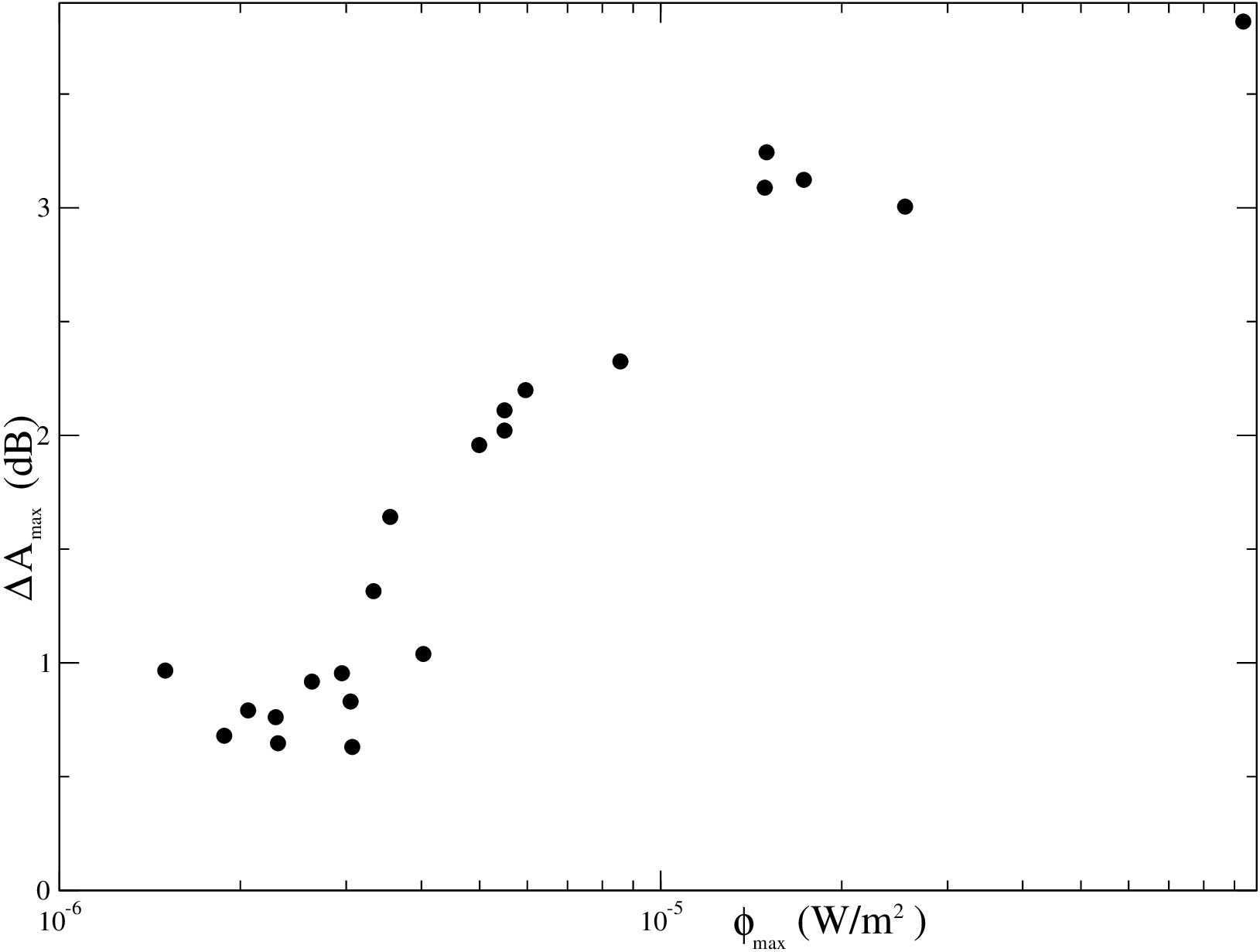}
\end{center}
\caption{Maximum perturbed VLF amplitude $\bigtriangleup A$$_{max}$ (in dB) is
 plotted as a function of corresponding peak flare flux $\phi_{max}$ (W m$^{-2}$)
 for all 22 flares.}
\end{figure}

\begin{table*}
\caption{Data sheet of all included flares\label{tbl-2}}
\begin{tabular}{@{}lllllllll@{}}
\tableline
$Date$& $time (s) $& $\phi_{max}$ $(W m^{-2})$& $Flare$ & $\bigtriangleup A_{max}$& $\chi (deg.)$& $\bigtriangleup t (s)$ &$N_{e,max}$ & $\alpha_{eff}$\\
$(yyyymmdd)$& $[IST]$& &  $class$& $(dB)$& & & $(m^{-3})$& $(m^{3} s^{-1})$\\
\tableline
20110121&35230&$3.33 \times 10^{-6}$ &$C3.33$ &1.315 &$27.0$ &$151$ &$5.19 \times 10^{9}$ &$9.155 \times 10^{-13}$\\
20110210&43560&$1.88 \times 10^{-6}$ &$C1.88$ &0.68 &$28.7$ &$353$ &$3.34 \times 10^{9}$ &$153.44 \times 10^{-13}$\\
20110210&44873&$2.06 \times 10^{-6}$ &$C2.06$ &0.791 &$32.4$ &$307$ &$3.63 \times 10^{9}$ &$23.94 \times 10^{-13}$\\
20110216&40526&$5.96 \times 10^{-6}$ &$C5.96$ &2.199 &$20.5$ &$222$ &$9.71 \times 10^{9}$ &$5.319 \times 10^{-13}$\\
20110218&37272&$4.03 \times 10^{-6}$ &$C4.03$ &1.039 &$18.0$ &$183$ &$4.38 \times 10^{9}$ &$27.67 \times 10^{-13}$\\
20110218&43365&$8.57 \times 10^{-6}$ &$C8.57$ &2.325 &$26.7$ &$147$ &$10.84 \times 10^{9}$ &$5.523 \times 10^{-13}$\\
20110219&45278&$2.63 \times 10^{-6}$ &$C2.63$ &0.918 &$32.4$ &$247$ &$3.86 \times 10^{9}$ &$23.197 \times 10^{-13}$\\
20110308&34140&$15.0 \times 10^{-6}$ &$M1.5$ &3.244 &$24.3$ &$121$ &$26.65 \times 10^{9}$ &$1.752 \times 10^{-13}$\\
20110308&38246&$5.5 \times 10^{-6}$ &$C5.5$ &2.021 &$14.2$ &$140$ &$8.97 \times 10^{9}$ &$5.182 \times 10^{-13}$\\
20110310&34052&$2.95 \times 10^{-6}$ &$C2.95$ &0.955 &$24.6$ &$264$ &$4.06 \times 10^{9}$ &$67.33 \times 10^{-13}$\\
20110311&36141&$5.5 \times 10^{-6}$ &$C5.5$ &2.11 &$17.9$ &$140$ &$9.89 \times 10^{9}$ &$4.4 \times 10^{-13}$\\
20110311&45175&$3.05 \times 10^{-6}$ &$C3.05$ &0.83 &$29.6$ &$181$ &$3.70 \times 10^{9}$ &$21.175 \times 10^{-13}$\\
20110414&39420&$4.99 \times 10^{-6}$ &$C4.99$ &1.957 &$14.3$ &$134$ &$8.60 \times 10^{9}$ &$5.282 \times 10^{-13}$\\
20110416&40071&$3.55 \times 10^{-6}$ &$C3.55$ &1.641 &$14.7$ &$150$ &$6.45 \times 10^{9}$ &$6.542 \times 10^{-13}$\\
20110607&43800&$25.5 \times 10^{-6}$ &$M2.55$ &3.005 &$28.8$ &$90$ &$18.36 \times 10^{9}$ &$5.929 \times 10^{-13}$\\
20110727&43760&$3.07 \times 10^{-6}$ &$C3.07$ &0.63 &$26.8$ &$179$ &$3.270 \times 10^{9}$ &$37.65 \times 10^{-13}$\\
20110728&36861&$2.29 \times 10^{-6}$ &$C2.29$ &0.761 &$23.2$ &$190$ &$3.48 \times 10^{9}$ &$14.855 \times 10^{-13}$\\
20110802&42542&$14.9 \times 10^{-6}$ &$M1.49$ &3.088 &$22.6$ &$128$ &$20.57 \times 10^{9}$ &$2.626 \times 10^{-13}$\\
20110803&36131&$17.3 \times 10^{-6}$ &$M1.73$ &3.123 &$23.8$ &$145$ &$23.71 \times 10^{9}$ &$2.283 \times 10^{-13}$\\
20110804&34017&$93.1 \times 10^{-6}$ &$M9.31$ &3.818 &$29.3$ &$74$ &$44.5 \times 10^{9}$ &$4.835 \times 10^{-13}$\\
20110817&35940&$2.31 \times 10^{-6}$ &$C2.31$ &0.647 &$22.1$ &$298$ &$3.82 \times 10^{9}$ &$29.61 \times 10^{-13}$\\
20110830&44009&$1.5 \times 10^{-6}$ &$C1.5$ &0.966 &$24.7$ &$241$ &$4.06 \times 10^{9}$ &$6.832 \times 10^{-13}$\\
\tableline
\end{tabular}
\end{table*}

\section{Effective recombination coefficient during solar flares}

\begin{figure}[h!]
\begin{center}
\includegraphics[height=6.3cm]{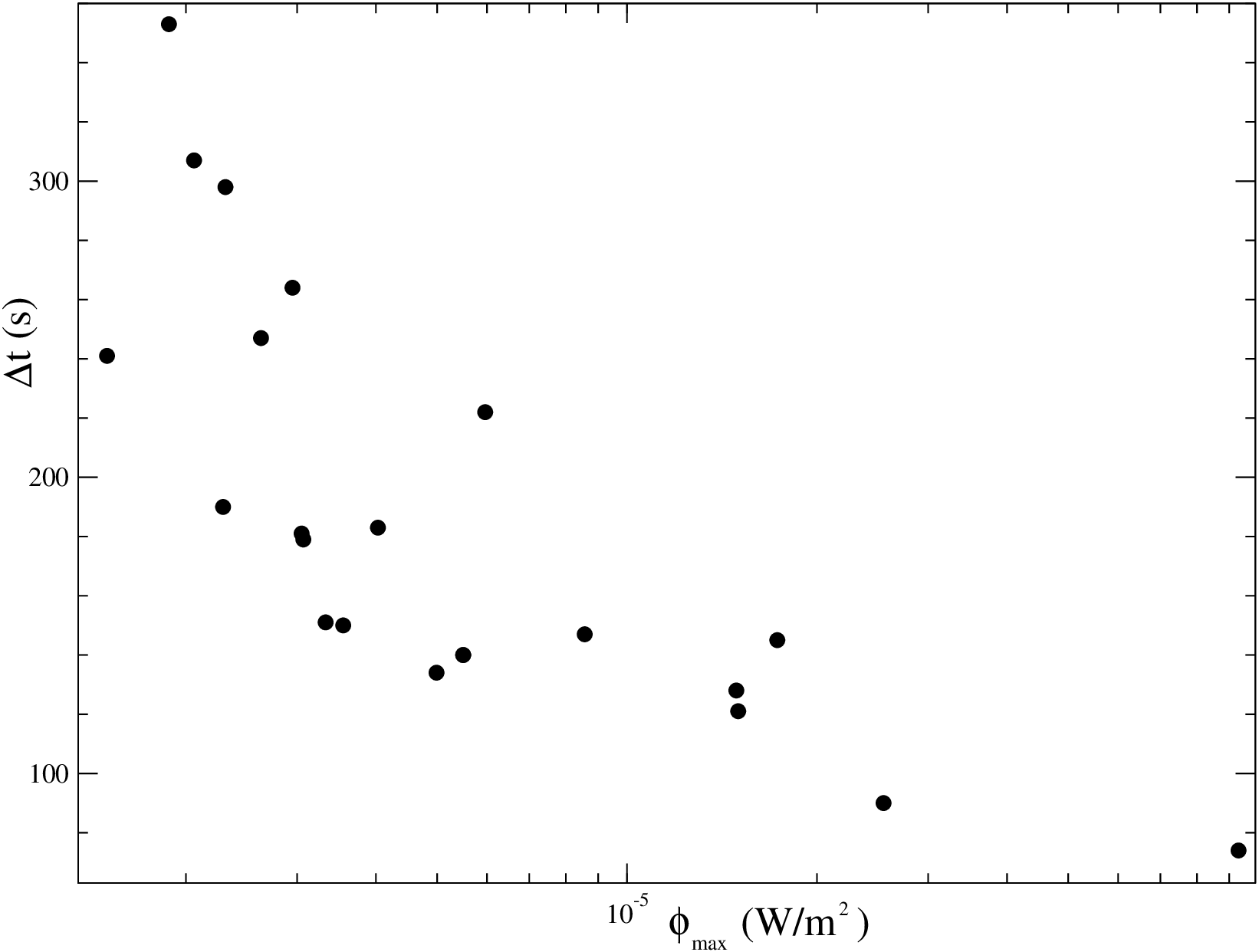}
\end{center}
\caption{VLF amplitude time delay ($\bigtriangleup t$) is plotted as a function
 of corresponding peak flare flux ($\phi_{max}$, W m$^{-2}$) for the same flares
 showed in Fig. 4 }
\end{figure}

\subsection{Theoretical understanding}

Our main aim in this Section is to obtain the variation of the {\it effective
recombination coefficient} ($\alpha_{eff}$) of D-region. The fundamental assumption of this
calculation is that the sub-ionospheric VLF signal quickly senses the changes in D-region electron
density ($N_e$) and modifies itself accordingly. Therefore, the VLF signal amplitude almost follows
temporal variation of $N_e$ even during flares (i.e., t$_{\bigtriangleup A_{max}}$ = 
t$_{N_{e,max}}$). Keeping total charge neutrality in mind, the D-region electron continuity eqn. can be
written as \citep{whit61,row70}, 
\begin{equation}
\frac{dN_e}{dt} = \frac{q(t)}{1 + \lambda} - \frac{N_e}{1 + \lambda}
\frac{d\lambda}{dt} - [\lambda^2 \alpha_i + \lambda (\alpha_i + \alpha_D) + \alpha_D] {N_e}^2 ,
\end{equation}
where, $\alpha_i$ is the ion-ion recombination coefficient, $\alpha_D$ is the dissociative
recombination coefficient and $\lambda$ is the ratio of negative ion and electron density.
The electron production rate of the D-region due to $\phi(t)$ during the flare is given by \citep{zig07}, 
\begin{equation}
q(t) = \frac{C\phi(t)}{eH} \cos \chi,
\end{equation}
where, $e = 2.71828$, $C = \rho^{-1}$ ($\rho$ is the amount of energy required to
create an electron-ion pair) and the scale height is given by \citep{mitra92}, 
\begin{equation}
H = \frac{k_b T}{gm_{avg}} ,
\end{equation}
where, $k_b$ is the Boltzman constant, $g$ is the gravitational acceleration, $m_{avg}$ is the mean
molecular mass and $T$ is the average temperature of lower ionosphere. According to
\citet{row70,mitra74}, $\lambda$ is a slowly varying function of time. So we set,
d$\lambda$/dt $\simeq$ 0. \citet{whit65,mitra74} reported that, $\lambda$ $\ll$ 1
for altitudes above $70$ km. So the eqn. (4) can be approximated as,  
\begin{equation}
\frac{dN_e}{dt} = q(t) - \alpha_{eff} {N_e}^2 ,
\end{equation}
where, {\it the effective recombination coefficient} as described in Sec. 1 is as follows, 
\begin{equation}
\alpha_{eff} = \lambda (\alpha_i + \alpha_D) + \alpha_D .
\end{equation}

Applying eqn. (7) near $t =$ $t_{N_{e,max}}$ and $t_{\phi_{max}}$, the {\it
effective recombination coefficient} would be \citep{zig07}, 
\begin{equation}
\alpha_{eff} = \frac{0.375}{\bigtriangleup t (N_{e,max} - q_{max}\bigtriangleup{t})}
\end{equation}
Comparing eqns. (5), (6) and (9) we get, 
\begin{equation}
\alpha_{eff} = \frac{0.375}{\bigtriangleup t (N_{e,max} - \frac{\phi_{max}
gm_{avg}\bigtriangleup{t}}{\rho ek_b T} \cos \chi)}
\end{equation}
Among the constant terms in eqn. (10), $m_{avg} = 4.8 \times 10^{-26}$ kg
 \citep{mitra92} and $\rho = 34$ eV \citep{whit65}. Using SROSS-C2 satellite data, \citet{sharma04} measured electron and ion temperature changes of ionosphere and they showed that electron temperature ($T_e$) changes from 1.3 to 1.9 times and ion temperature ($T_i$) changes from 1.2 to 1.4 times respectively during flares in comparison with normal days. Thus the change of $\alpha_{eff}$ due to these $T_e$, $T_i$ changes is less enough compared to other ion-chemical effect induced changes. So for simplicity of the model, we justifiably assumed a thermal equilibrium and took $T = 210$ K \citep{sch11} to perform the entire calculation.
 Since only solar flares which occurred close to the mid-day have been considered in this
 analysis to eliminate dependence on $Z$, we take mean $\cos$ $Z = 0.90$
 for our calculation. This may be justified because for the flares included here, the 
zenith angle varies between $33^\circ$ to $14^\circ$ (all $Z$ values corresponding to 
each flare peak time are in Table 1). Our simulation 
procedure to obtain $N_{e,max}$ has been discussed in the next section.

\subsection{Simulation technics using LWPC}

Long Wave Propagation Model (LWPM) is the default propagation model of the lower
ionosphere \citep{ferg98}. It has exponential increase in $N_e$ and conductivity.
A sharpness factor ($\beta$) and effective reflection height ($h^\prime$) define this ionosphere model \citep{wait64}. $h^\prime$ = $74$ km and $\beta$ = $0.3$ km$^{-1}$ are
constant values being assumed to define daytime unperturbed ionosphere, even constant for
entire VLF range. Here we did the simulation of lower ionosphere over the single T$_x$-R$_x$ propagation path from NWC transmitter to IERC, Sitapur receiving station. As we mentioned in Sec. 4.1 that in $\alpha_{eff}$ analysis to avoid $Z$ dependence, we only accommodated those flares which occurred when the sun was close to local zenith (all $Z$ averaged over T$_x$-R$_x$ path are mentioned in Table 1) w.r.t. VLF signal propagation path. So we assumed that flare induced perturbation of the reflection height for entire horizontal path segments are same. 

We used Range Exponential model to handle the ionosphere perturbed due to
flares. First, we have added (see Eqn. 2) the $\bigtriangleup A$$_{max}$ with unperturbed ionosphere VLF amplitude
($A_{lwpc}$) corresponding to $h^\prime$ = $74$ km and $\beta$ = $0.3$ km$^{-1}$ \citep{pal10,pal12a,pal12b}.
The unperturbed D-region electron density is calculated as, $N_{e,unperturbed} = 2.2 \times 10^8 m^{-3}$.
To authenticate $N_{e,unperturbed}$, we obtain the same unperturbed value (averaged over entire propagation path) from IRI-model of NASA as, $(N_{e,unperturbed})_{IRI} = 4.4 \times 10^8 m^{-3}$.
There is little disagreement because LWPC follows Wait's 2-parameter lower ionospheric approximate model instead of exact realistic model. This much of disagreement of those values is justified and would not affect the physical situation as flare induced $N_e$ is of higher magnitude than these by few orders.

Now we run the program to simulate Wait's parameters corresponding to solar flare perturbed
ionosphere VLF amplitude ($A_{lwpc}$ + $\bigtriangleup A$$_{max}$) and this process is
repeated for all $22$ flares. 
\begin{figure}[h!]
\begin{center}
\includegraphics[height=6.5cm]{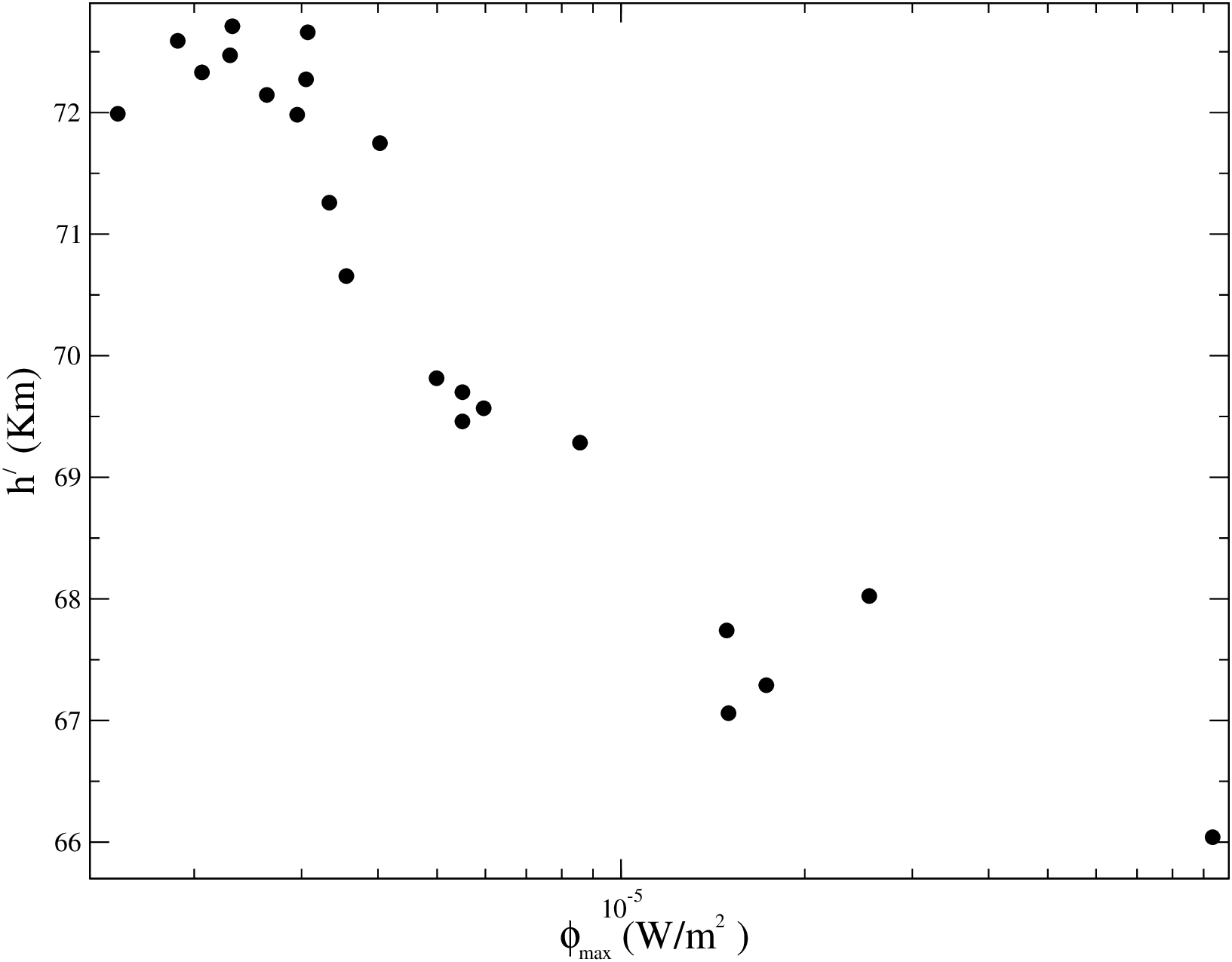}
\end{center}
\caption{Simulated effective reflection height ($h^\prime$) (km) is plotted with
corresponding peak flare flux ($\phi_{max}$) (W m$^{-2}$)}
\end{figure}

\begin{figure}[h!]
\begin{center}
\includegraphics[height=6.5cm]{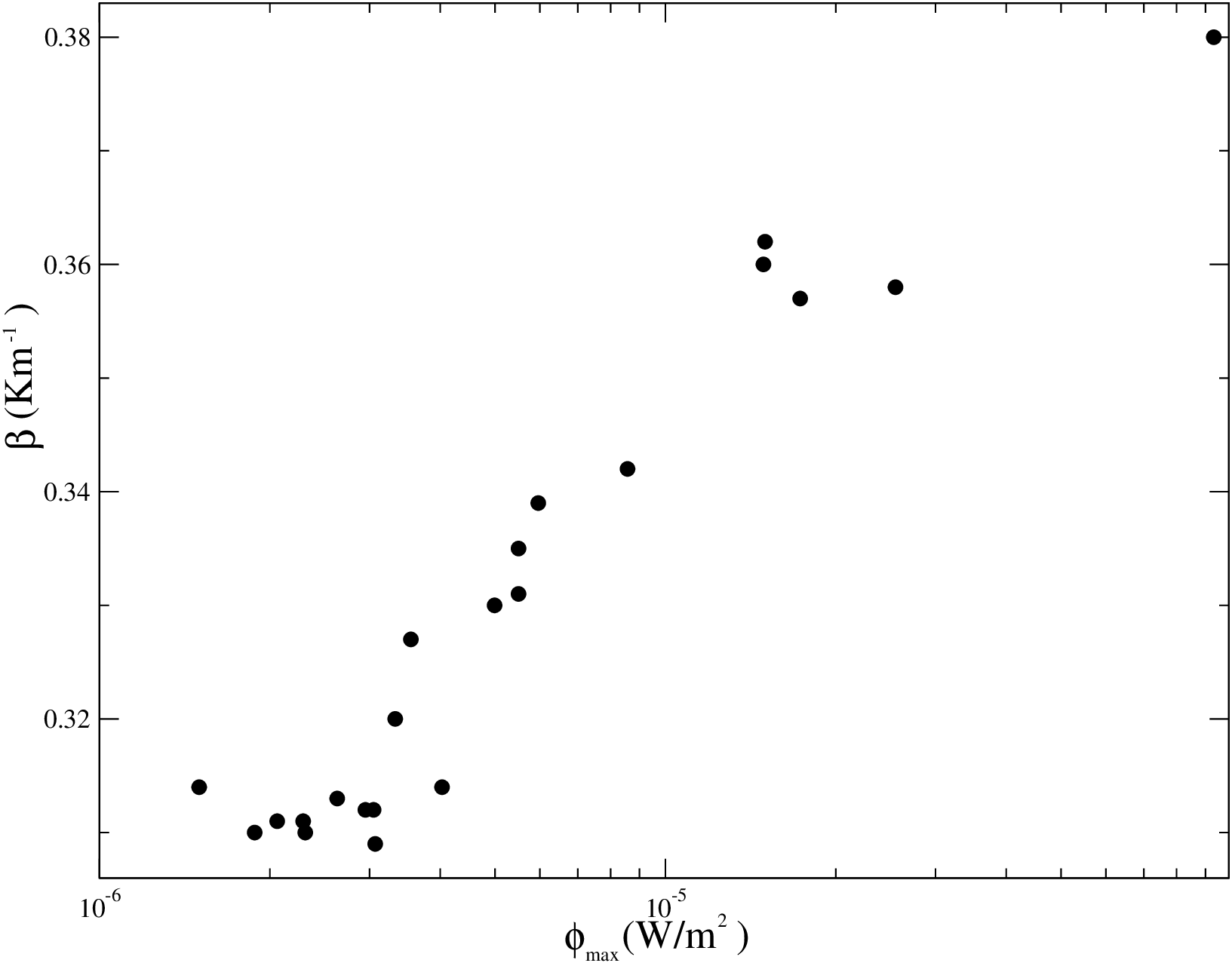}
\end{center}
\caption{Simulated sharpness factor ($\beta$) (km$^{-1}$) is plotted with
corresponding peak flare flux ($\phi_{max}$) (W m$^{-2}$)}
\end{figure}
In Figs. 6 and 7 these h$^\prime$ and $\beta$ are plotted with corresponding
$\phi_{max}$ respectively. Now using Wait's formula \citep{wait64,pal10,pal12a,pal12b}, 

\begin{equation}
N_{e,max} \propto e^{-\beta h^\prime} e^{(\beta - \beta_0)h}
\end{equation}
where, $\beta_0$ = $0.15$ km$^{-1}$, we get electron density at height $h$. We repeated the entire LWPC simulation for several complex ion density profile values and we noted hardly any changes for VLF signal amplitude perturbation results. Some physical reasons in support of it, have been discussed in Sec. 1.

\subsection{Results}

In Sec. 4, we analysed $22$ solar flares having $Z$ within $\sim$ $15^\circ$ to $30^\circ$. We got the expression for
$\alpha_{eff}$ as eqn. (10) and from LWPC simulation we estimated $N_{e,max}$ for each
flare (eqn. 11). In Fig. 8 the LWPC simulated $N_{e,max}$ (at height $h = 74$ km) 
is plotted with corresponding $\phi_{max}$. Also for verification of our LWPC simulation
 results, we again calculated $N_{e,max}$ at $h = 74.1$ km height using $\alpha_{eff}$
 values taken from \citet{zig07}. Though the simulated $N_{e,max}$ values are a bit 
underestimated w.r.t. the values adapted from literature but the general agreement
 is good by keeping our measurement errors in view.
\begin{figure}[h!]
\begin{center}
\includegraphics[height=6.4cm]{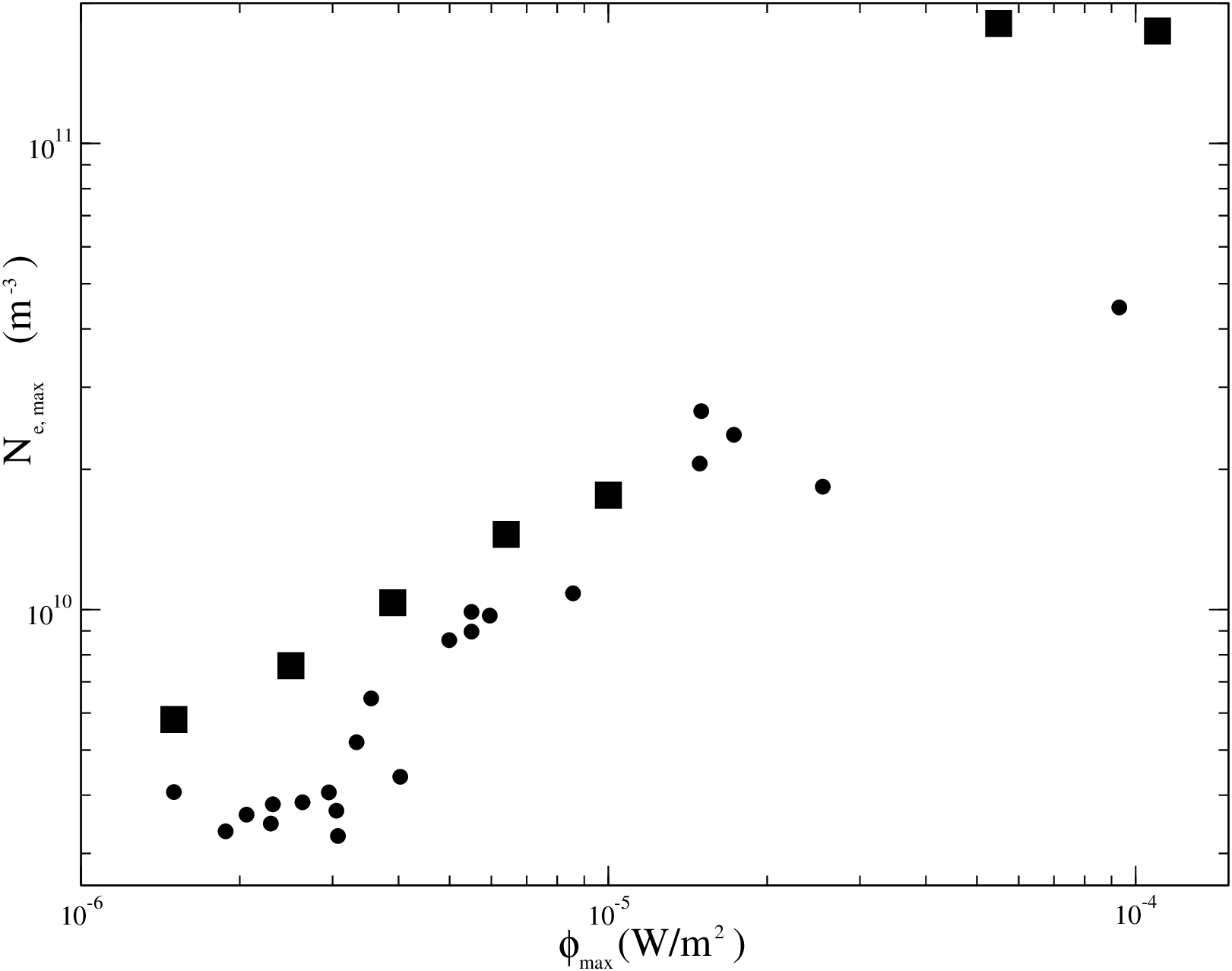}
\end{center}
\caption{(a) Simulated D-region electron density (circles) ($N_{e,max}$) (m$^{-3}$) at $h = 74$ km 
is plotted with corresponding peak flare flux ($\phi_{max}$) (W m$^{-2}$) (b) Also $N_{e,max}$ (squares) calculated at $h = 74.1$ km using $\alpha_{eff}$ values taken from \citet{zig07}}
\end{figure}
\begin{figure}[h!]
\begin{center}
\includegraphics[height=6.4cm]{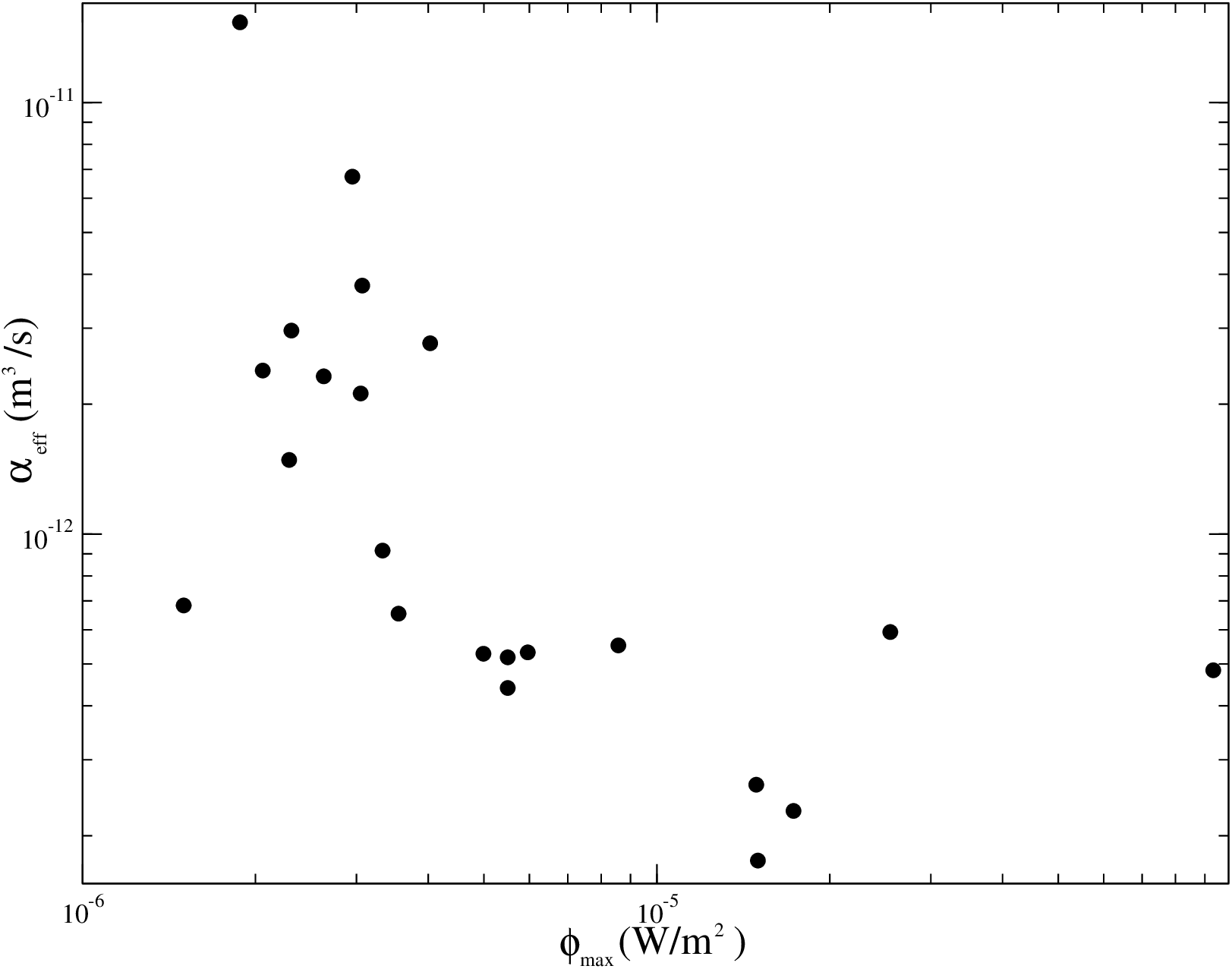}
\end{center}
\caption{{\it effective reflection coefficient} ($\alpha_{eff}$) ($m^3$s$^{-1}$) is
plotted with corresponding peak flare flux ($\phi_{max}$) (W m$^{-2}$)}
\end{figure}
Now substituting each $N_{e,max}$ to eqn. (10), $\alpha_{eff}$ can be calculated (see Table 1).
In Fig. 9, $\alpha_{eff}$ versus $\phi_{max}$ has been shown, the values of
$\alpha_{eff}$ goes from $\sim$ $10^{-13}$ $m^3$s$^{-1}$ to more than  $10^{-11}$
m$^3$s$^{-1}$ for $\sim$ C1.0 to M9.0 classes of flares. We see that $\alpha_{eff}$ and $\phi_{max}$ are
generally correlated with some exceptions. Those may have appeared due to some 
zenith angle ($Z$) variation in the observation. The calculated values of $\alpha_{eff}$
for low latitude D-region ionosphere generally agree with the results of
\citet{par65,bal91,zig07}. \citet{fehsen69,mitra72,mitra74} reported that increase in
solar flux intensity transforms the water-cluster ions to molecular ions. Though that
 process in dominant near mesopause region, but we report that at $74$ km in D-region 
that transformation occurs at significant level and hence
$\alpha_{eff}$ is reduced. Our results support this observation.

\section{Effects of solar zenith angle variation on flare perturbed D-region}
\subsection{Zenith angle effects on time delay during C-class flares}

In the section we would discuss 
the variation of ionospheric time delay ($\bigtriangleup t$) during flare 
with $Z$ over the NWC-IERC signal propagation path. We calculate the 
average zenith angle ($Z$) by taking mean of the zenith 
angle values at every $10 km$ interval on the propagation path at the peak time of the 
flare. We did it to incorporate the different solar radiation 
inclinations at different path segments. We computed this for the $22$ C-class 
flares which are from C2 to C7. One of the reasons to confine our discussion for a single C-class is 
this: the $\bigtriangleup t$ has dependence on $N_{e,max}$ and 
hence on $\phi_{max}$ according to \citet{mitra74}. 
If we mix various classes of flares, such a dependence might shadow the effect we are after.
Thus we concentrated on flares of same class occurring at various times of the 
day. We could do similar analysis for other classes also, but the their numbers were 
too few to establish a correlation beyond any reasonable doubt.

\begin{figure}[h!]
\begin{center}
\includegraphics[height=6.6cm]{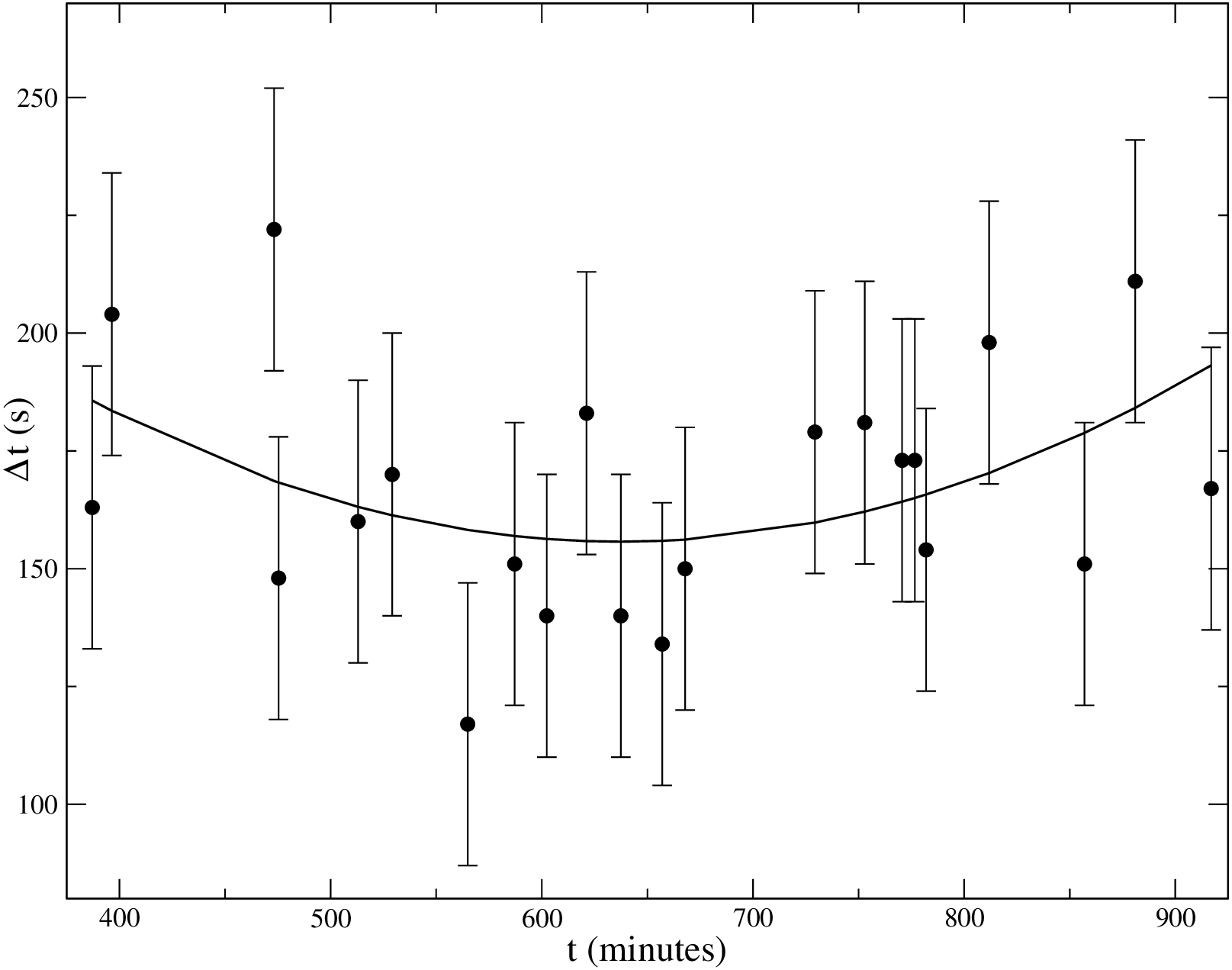}%
\end{center}
\caption{The time delay ($\bigtriangleup t$) (s) of the ionosphere for each 
C-class solar flare is plotted as a function of their occurrence time 
t (mins) of the corresponding solar flare. The solid line is the fitted 
mean curve for those data points}
\end{figure}

In Fig. 10, the $\bigtriangleup t$s are plotted against their 
occurrence time ($t$) [IST] and fitted curve has been drawn to 
actually show the variation with diurnal time and hence 
with $Z$. It is interesting that the mean curve closely 
follow the typical diurnal $Z$ variation. The $Z$ starts decreasing 
after sunrise. It is minimum during the mid-day when the sun is close to 
zenith at the top and gradually increases towards the sunset. 
The errorbars of $\bigtriangleup t$ represent the resolution of the 
recorded VLF amplitude data and the available online X-ray flux. \citet{le07,le12} have shown  
that the $Z$-dependence of the upper ionosphere using       
TEC measurements and besides we show the $Z$ dependence of lower
ionospheric response.
\begin{figure}[h!]
\begin{center}
\includegraphics[height=6.5cm]{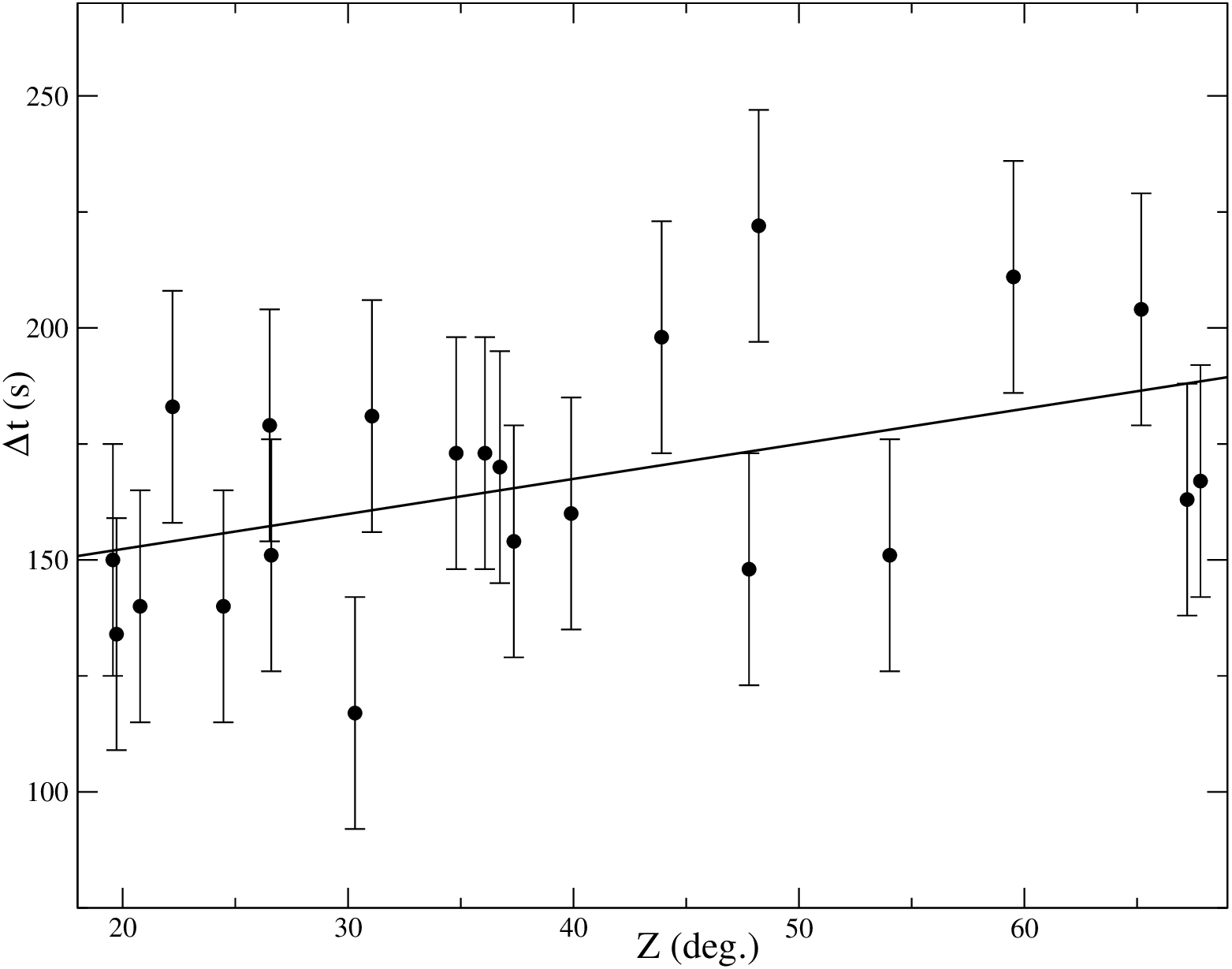}%
\end{center}
\caption{The time delay ($\bigtriangleup t$) (s) of the ionosphere for each 
C-class solar flare is plotted as a function of $Z$ at their occurrence 
time of the corresponding solar flare. The solid line is the fitted 
straight line for those data points}
\end{figure}
 For further analysis, we plot the 
$\bigtriangleup t$ directly with corresponding $Z$ (Fig. 11)
and fitted them with a straight line to draw direct correlation. In the equation of the 
straight line,
\begin{equation}
{\bigtriangleup t} = a_1 Z + a_2,
\end{equation} 
The fitted values are $a_1 = 0.7556$ deg$^{-1}$ s and $a_2 = 137.24$ s. The goodness of the
fit established as the $reduced$-$\chi^2 = 0.92$ (No. of data points $N = 22$ 
and fitted using, $P=2$ parameters (see eqn. 12), so the available degrees of freedom for 
this $\chi^2$ - test is, $N - P = 22 - 2 = 20$). We see that 
the $reduced$-$\chi^2$ is close to unity and thus correlation is very good.
From a physical point of view this correlation can be explained in the following way: 
the residual degree of ionization and free electron density ($N_e$) of lower ionosphere 
is mostly governed by $Z$. During solar flares, bombardment of higher energetic 
X-ray photons on ionospheric bed leads to evolution and excitation but the $Z$-dependence of the
flux remains. 
\subsection{Zenith angle dependence of correlation between $\phi_{max}$ and $\bigtriangleup t$}
\subsubsection{Grouping and analysis of solar flares}

\begin{table*}
\caption{Details of time-bin\label{tbl-2}}
\begin{tabular}{@{}crrrrrrrrr@{}}
\tableline
  & Time range (s)& C-type & M-type & X-type & $a$ $(s)$ & $b$ & $Red$-$\chi^2$ & $ f= N - P$ & $\sigma (deg.)$ \\
\tableline
DT1 &22000-28000 &6  &4 &1 &347.0 &0.39  &0.923 &9  &3.231\\
DT2 &28000-34000 &14 &3 &0 &247.1 &0.23  &1.039 &15 &3.084\\
DT3 &34000-40000 &8  &3 &0 &311.3 &0.35  &1.324 &9  &6.535\\
DT4 &40000-46000 &8  &2 &0 &346.8 &0.40  &1.532 &8  &13.698\\
DT5 &46000-52000 &11 &4 &1 &219.4 &0.29  &1.663 &14 &12.629\\
DT6 &52000-58000 &12 &1 &0 &238.6 &0.30  &1.725 &11 &12.26\\
\tableline
\end{tabular}
\end{table*}

We binned 78 flares of C, M and X classes into six separate 
$100$ $mins$ sized time-bins namely, $DT1$, $DT2$, $DT3$, $DT4$, $DT5$ and $DT6$. 
The range of time-bins and details of flares present in those bins
are given in Table 2. We chose the size of bins in a way that 
bins are sufficiently small so that one could assume $Z$ to be constant in each bin,
while they are sufficiently big, so that there are statistically 
significant number of flares in each bin.
Firstly we neglected the $Z$ variation within a given bin and secondly
for validity of our statistical results we accommodated sufficient number of
flare cases in each bin from available VLF data.

Time delay ($\bigtriangleup t$) is defined by eqn. (1) is similar to the time dilation
parameters defined by Appleton (1953); Valnicek and Ranzinger (1972); Zigman (2007) etc. We first fitted the X-ray and VLF data 
analytically before obtaining $\bigtriangleup t$ (Sec. 3). In the sec. 5.1, we sorted only C-class flares and showed that, $\bigtriangleup t = f_1(Z)$, i.e.,
$\bigtriangleup t$s for these flares depend on mean $Z$ computed over the signal propagation path. 

While calculating $\alpha_{eff}$ in sec. 4, we showed that, $\bigtriangleup t = f_2(\phi_{max})$, i.e.,
 $\bigtriangleup t$ significantly depends of the corresponding peak flare flux 
$\phi_{max}$. In order to demonstrate, $\bigtriangleup t = f(Z, \phi_{max})$, 
i.e., the dependence of  $\bigtriangleup t$ on $\phi_{max}$ and $Z$ simultaneously,
we introduce this new approach.
Observing the nature of dependence we assume an empirical relation between the $\bigtriangleup t$ and $\phi_{max}$ for the flares in each bin:
\begin{equation}
{\bigtriangleup t} = a e^{-b (\log_{10} \phi_{max})},
\end{equation}
where, $a$ and $b$ are real parameters. Then we fitted data points of each bins separately 
using non-linear least square fitting technique using $Gnuplot$ software and $a$ and $b$ 
are the parameter values (see Table 2) corresponding to best fitting of eqn. 13 
and calculated $reduced$-$\chi^2$ for every fit to test the `goodness of fit'. The $reduced$-$0\chi^2$ for $j$-th bin is,
\begin{equation}
{reduced-\chi^2_j} = {\frac{1}{f}}{\sum_{\substack{DT}}{\frac{(\bigtriangleup t_{obs}-\bigtriangleup t_{em})^2}{\sigma_{err}^2}}},
\end{equation}
where, $\bigtriangleup t_{obs}$ is experimentally observed time delay, $\bigtriangleup t_{em}$ is the time delay calculated using eqn. 14, $\sigma_{err}$ is the size of the errorbar, which comes from the limitations of the measuring instrument. In our case $\sigma_{err} = 30 s$. Now the degree of freedom is defined as,
\begin{equation}
f = N - P
\end{equation}
where, $N$ is the number accommodated flares within $j$-th time-bin and $P$ is number of model parameters used to fit $N$ number of data points and here $P = 2$ from eqn. 13. Our main objectives in this paper is to check the nature of the
$\bigtriangleup t$ at different $Z$. 
This is because, the residual degree of ionization (due to Lyman-$\alpha$ and soft X-ray) 
of lower ionosphere is mainly governed by $Z$. In Fig. 12, we plot the 
variation of $Z$ along $T_x$-$R_x$. Six different graphs represent
typical $Z$ variation at midway of each time bin. 
\begin{figure}[h!]
\begin{center}
\includegraphics[height=6.5cm]{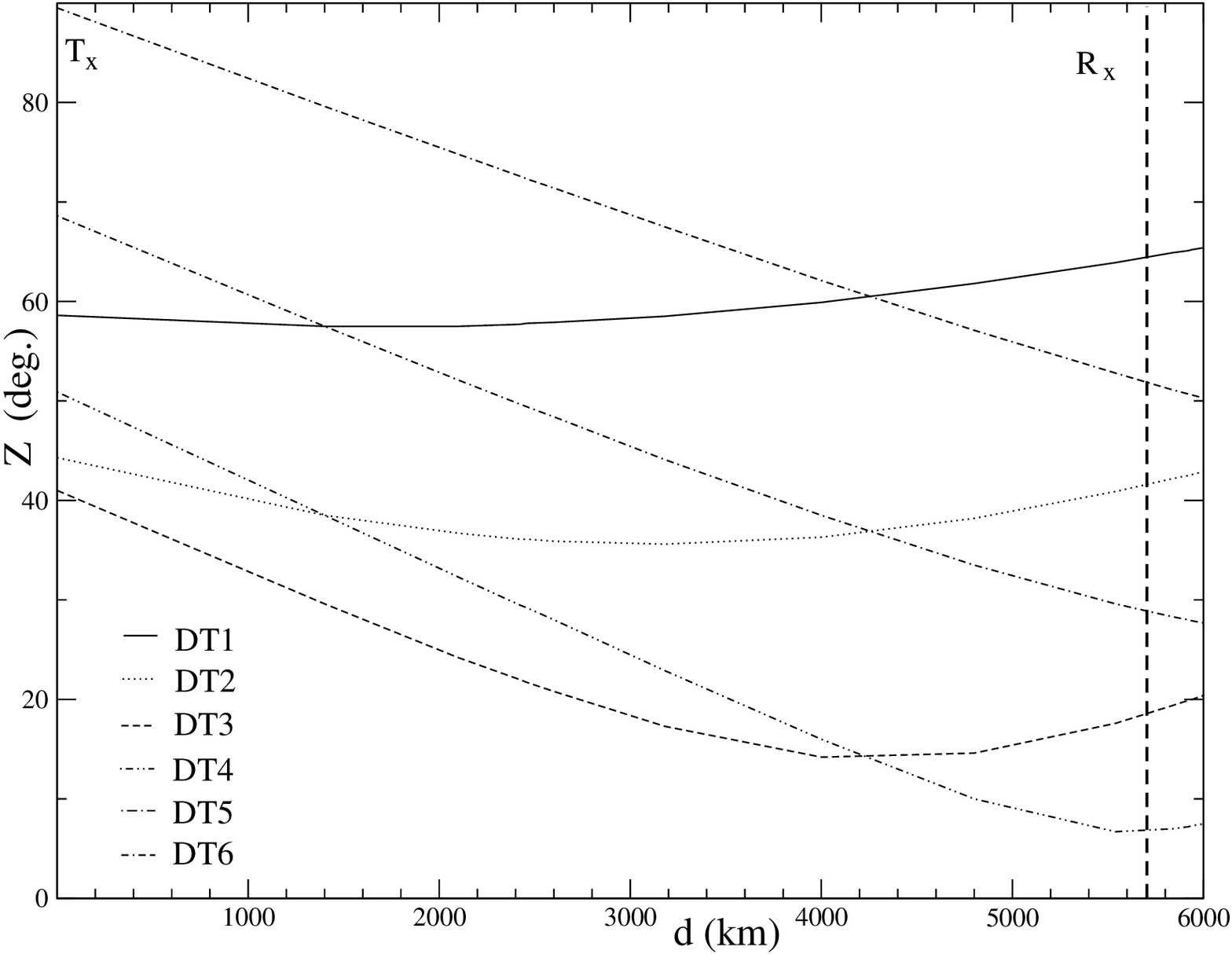}
\end{center}
\caption{Variation of mean zenith angle ($Z$) (deg.) as a function of $T_x$-$R_x$ 
path at middle of time-bins mentioned in Table 2. $T_x$ is at $d=0 km$ and $R_x$ 
is at $d=5691 kms$ }
\end{figure}
Along the entire path, $Z$ varies considerably for some time bin (e.g. $DT3$, $DT4$, $DT5$, $DT6$ mainly), and thus it is not possible to
assume $Z$ to be a constant. One way to quantify $Z$ for each time bin, would be to 
measure `how bad' the assumption of constant $Z$ is. In other words, we would be interested in obtaining 
the standard deviation ($\sigma$) of $Z$ for each time bins. This $\sigma$ will thus be an estimate 
of variation of residual ionization due to the flare alone along $T_x$-$R_x$ path. 
In the next Section, we would test our hypothesis that if $\sigma$ is large, i.e., the 
variation of ionization along the path is significant, then the $reduced$-$\chi^2$ 
is also going to be large, i.e., the correlation between $\bigtriangleup t$ and 
$\phi_{max}$ is also loose. The opposite should also be correct.

\subsubsection{Results}
We have analysed a total of 78 flares (Table 2) and for each
flare, the VLF peak appeared regularly after an X-ray. Hence for all the cases under this investigation we got
$\bigtriangleup t\textgreater 0$. Grouping techniques and analysing 
procedures of flares are also discussed in earlier section.
\begin{figure}[h!]
\begin{center}
\includegraphics[height=6.5cm]{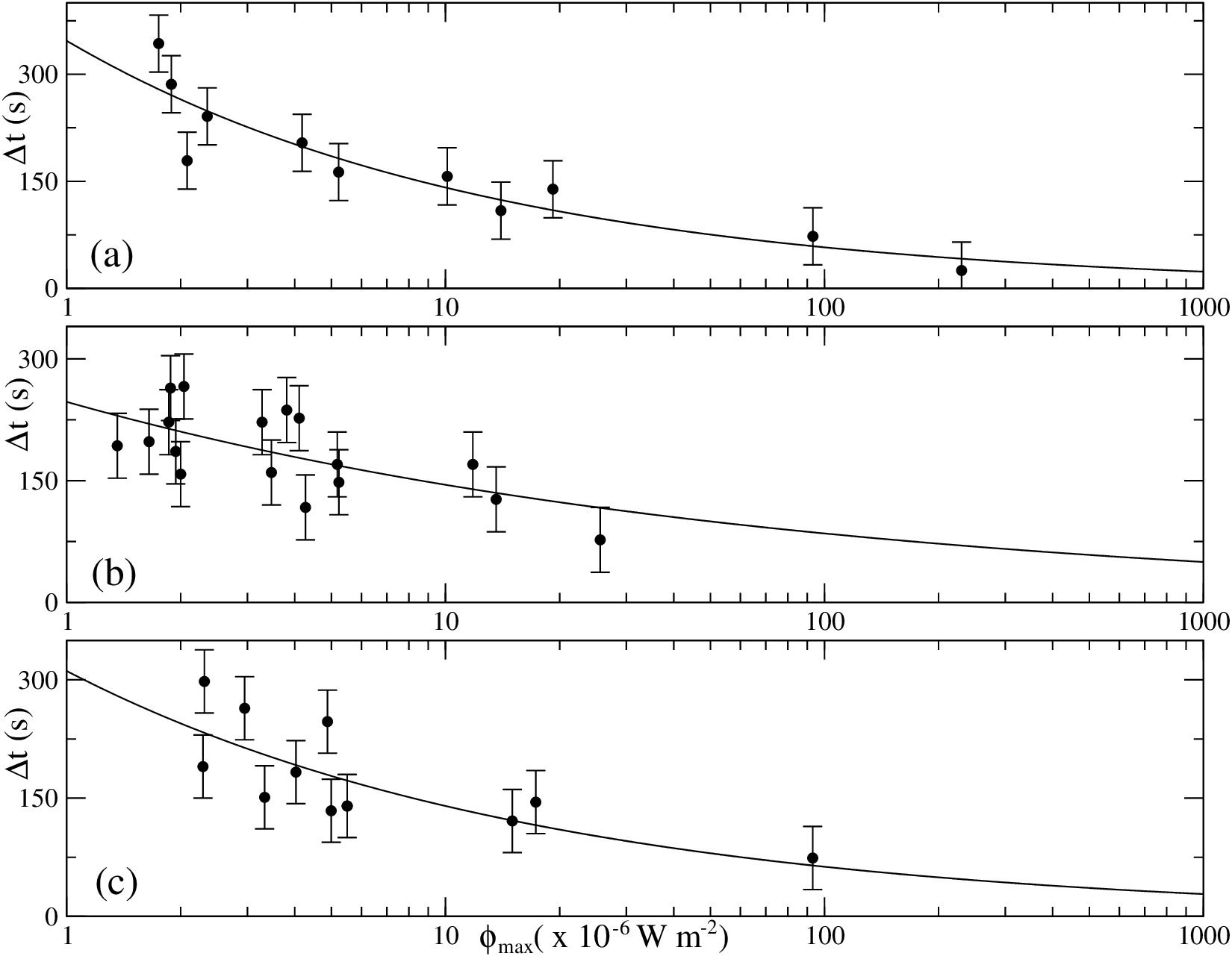}
\end{center}
\caption{The time delay ($\bigtriangleup t(s)$) for each solar flare has been
 plotted as a function of peak flare flux ($\phi_{max}$ in  W m$^{-2}$). The solid
 line is the fitted empirical function (eqn. 13). The time bins are (a)$DT1$, (b)$DT2$ and (c)$DT3$} 
\end{figure}

\begin{figure}[h!]
\begin{center}
\includegraphics[height=6.5cm]{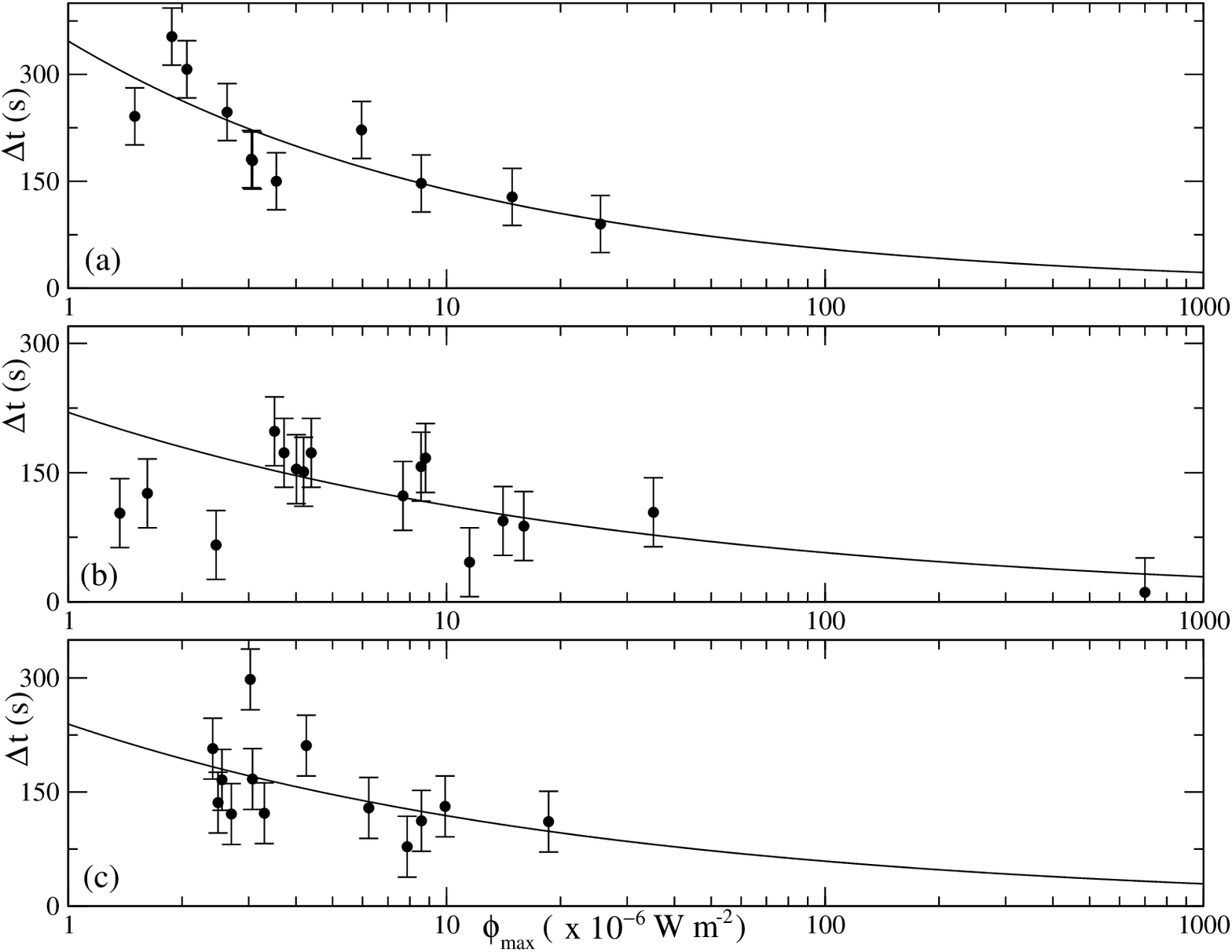}
\end{center}
\caption{The same plot of time delay ($\bigtriangleup t$) as in Fig. 13. 
The time bins are (a) $DT4$, (b) $DT5$ and (c) $DT6$} 
\end{figure}

All the flare events are plotted bin wise in Fig. 13 and Fig. 14. Error bars
reflect resolutions of available GOES-15 X-ray data from NOAA archive and IERC 
recorded VLF data. First, we note the decreasing tendency of $\bigtriangleup t$
with increasing $\phi_{max}$, which we already reported during $\alpha_{eff}$ for limited
number of flares within small span of $Z$. After fitting them with the empirical function (eqn. 13) the fitting
parameters $a$ and $b$ are evaluated (Table 2). Thereafter we calculate
$reduced$-$\chi^2$ for each fit to estimate the `goodness of fit' (the degrees of freedom,
$f$ corresponding to those fitting are given in Table 2). The $reduced$-$\chi^2$ 
physically represents the dominance of $\phi_{max}$ on $\bigtriangleup$t
during the flares at an effective ionization level of the D-region governed by $Z$. 
Six different time bins represents discrete residual ionized states 
caused by different path averaged $Z$ values. The $reduced$-$\chi^2$ values are around `unity'
which means that the correlation is generally very good. However, as we progress from 
$DT1$ to $DT6$, we see that the $reduced$-$\chi^2$ increases. Thus the fitting 
becomes poorer and control of $\phi_{max}$ over $\bigtriangleup t$ weakens gradually.

So far we discussed the physical property of $\bigtriangleup t$. Possible physical
explanation of this, at least for the low-latitude, trans-equatorial, medium length 
signal propagation path (NWC-IERC) can be understood by the following exercise. 
In Fig. 12, the variation of $Z$ over $T_x$-$R_x$ path at mid-point of the time 
bins is very significant. This $Z$ variation and hence the variation of residual 
ionization level of D-region over the propagation path increases monotonically from $DT1$ to $DT6$.
The standard deviation ($\sigma$) of $Z$ denotes the spread of it over the path (Fig. 15).

Our findings indicate that the gradient of residual ionization level over $T_x$-$R_x$ path is 
the key determining factor of $\bigtriangleup t$ for different classes of solar X-ray 
flares occurred at different times. 
\begin{figure}[h!]
\begin{center}
\includegraphics[height=6.3cm]{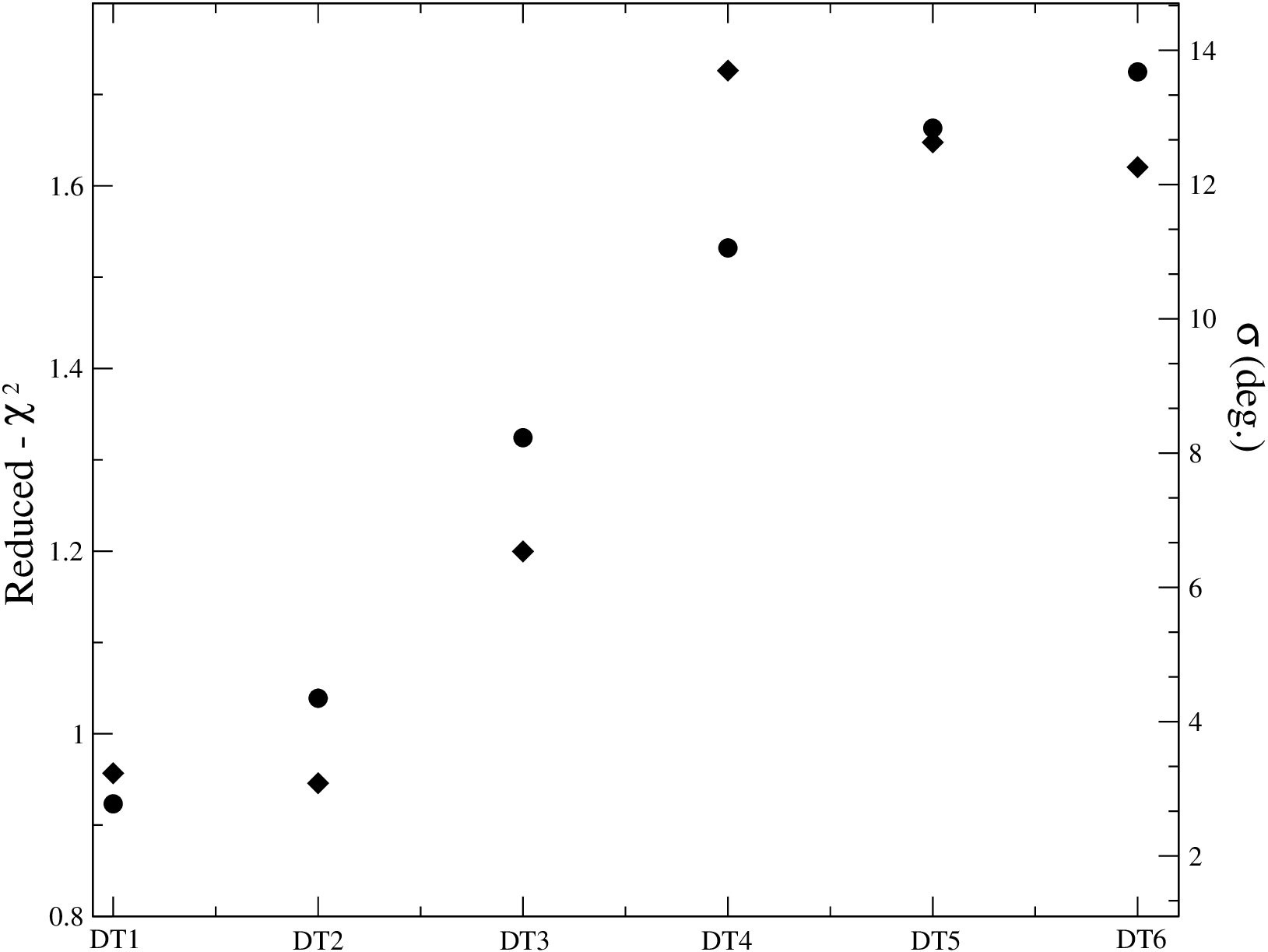}
\end{center}
\caption{The $reduced$-$\chi^2$ (circle) and standard deviation ($\sigma$) 
(diamond) of the zenith angle ($Z$) along $T_x$-$R_x$ path as functions of $DT1$, $DT2$, $DT3$ etc.} 
\end{figure}
\section{Conclusions} 

We monitored the X-ray data of GOES-15 and VLF data from NWC/19.8 kHz station as received at the IERC
station by SoftPAL receiver. These data are analysed in this paper. The data acquisition period 
is Jan-Sep 2011. For $\alpha_{eff}$ calculation, we analyse VLF data of 22 flares which occur near the local mid-day in order to 
eliminate the effects of the $Z$. The signals were from NWC/19.8 kHz transmitter
and are recorded at IERC/ICSP, Sitapur, India. For all the flares presented 
in this paper, we estimates $\bigtriangleup A$, $\bigtriangleup t$ 
$\textgreater$ 0. The $\bigtriangleup t$ is the important parameter 
in this analysis. From Fig. 5, we find a decreasing 
nature of $\bigtriangleup t$ with increasing $\phi_{max}$.
From this result, we can verify the \citet{appl53} relation regarding $\bigtriangleup t$.
Most importantly we calculated the effective recombination coefficient
($\alpha_{eff}$) at the peak of these flares using a
coupled theoretical model which consists of electron continuity theory
and LWPC simulation (Fig. 9). The values of $\alpha_{eff}$ at 74 km
altitude are generally in agreement with earlier results reported by
several other workers. The zenith angle dependence and latitude-longitude
dependence have not been studied this section.  

While checking direct connection between $Z$ and $\bigtriangleup t$, we concentrated on a very narrow range of flare energies (C2-C7) in order that
the dependence of $\bigtriangleup$t on $\phi_{max}$ may be ignored. The most important result 
we reported here is the linear correlation between the solar zenith angle ($Z$) averaged over the propagation path 
at the flare time and the time delay $\bigtriangleup$t between $FPT$ and $APT$ for that flare. 
The fit is excellent as the reduced $\chi^2$ is found to be close to unity. We believe that 
when we narrow our choice of flare energies even further, the correlation would be even tighter. This
and a similar study for M and X-class flares would be taken up when we have data for a sufficient
number of flares.

Now we moved towards a broader approach and we successfully analysed 78 flares of all classes. 
We generally find that the time delay of the VLF amplitude $\bigtriangleup t$
is anti-correlated with the flare energy flux $\phi_{max}$. However, the relationship is not 
equally good. From the $reduced$-$\chi^2$ of the fit between the 
 $\bigtriangleup t$ and $\phi_{max}$, we note that $reduced-\chi^2$ is low 
in the early hours of the day and it worsens as the time of the occurrence of 
flares progresses. To understand this behaviour we also plot the variation of zenithal angle of 
the sun $Z$ along the propagation path in different time slots. Most interestingly we find that the 
standard deviation of $Z$ in each slot, when plotted against the time slots, behave similarly as the 
variation of $reduced-\chi^2$ (Fig. 15). Though it is notable that $reduced-\chi^2$ is not following the $\sigma$ for each of those $DTs$. But the over all increasing tendencies of those parameters with $DTs$ are comparable. This indicates that the dispersion in the $\bigtriangleup t$ vs.
$\phi_{max}$ relationship primarily depends on how dispersed the ionization is along the 
propagation path and it happens specifically on the earlier times of the day i.e. at $DT1$, $DT2$, $DT3$ and $DT4$. But for $DT5$ and $DT6$ the $reduced-\chi^2$ and $\sigma$ behaves in slightly opposite manner. Dominant recombination processes in later part of the day may be the possible reason for that. In future we would verify this observation for more flare events.
 Moreover, confirmation of this observation for other $T_x$-$R_x$ pairs would
to make our result statistically more significant. We will also study the effect
for a larger number of propagation paths. This will be reported elsewhere.

\section{Acknowledgement}

We thank S. Pal and S. K. Mondal for discussions and ICSP for allowing us to use the IERC data. Tamal Basak acknowledges a CSIR
Fellowship for support.

\end{document}